\documentclass[prb,twocolumn,showkeys,nofootinbib,preprintnumbers,longbibliography,aps,superscriptaddress
]{revtex4-2}

\usepackage{amsmath,amssymb,ifpdf,url,mathrsfs,slashed,multirow,tabularx,type1cm}
\usepackage{subfigure}
\usepackage[dvipdfmx]{graphicx}
\usepackage{color}
\usepackage{braket}
\usepackage{bm}
\usepackage[toc,page]{appendix}
\usepackage{comment} 
\usepackage[normalem]{ulem}
\usepackage{here}

\usepackage{ascmac}
\usepackage{enumerate}

\allowdisplaybreaks
 \definecolor{BlueViolet}{rgb}{0.2, 0.00, 0.7}
\definecolor{Blue}{rgb}{0.15, 0.00, 0.9}
\usepackage[
colorlinks=true,linkcolor=Blue,citecolor=Blue,
urlcolor=BlueViolet,breaklinks=true]{hyperref}

\newcommand{\Slash}[1]{{\ooalign{\hfil \hspace*{-5pt}~#1\hfil\crcr\raise.167ex\hbox{/}}}}

\def\({\left(}
\def\){\right)}
\def\<{\langle}
\def\>{\rangle}

\newcommand{\matl}{\left( \begin{array}}
\newcommand{\matr}{\end{array} \right)}

\def\beq#1\eeq{\begin{align}#1\end{align}}

\newcommand{\du}[1]{\textcolor{blue}{#1}}

\begin{document}
\preprint{KEK--TH--2546}

\title{
Quantum resonance viewed as weak measurement
}

\author{Daiki Ueda} \email{uedad@post.kek.jp} 
\affiliation{KEK Theory Center, IPNS, KEK, Tsukuba 305-0801, Japan.}
\affiliation{Physics Department, Technion \text{--} Israel Institute of Technology,
Technion city, Haifa 3200003, Israel.}

\author{Izumi Tsutsui}
\email{itsutsui@post.kek.jp}
\affiliation{Department of Physics, College of Science and Technology, Nihon University,
Tokyo 101-8308, Japan.}
\affiliation{KEK Theory Center, IPNS, KEK, Tsukuba 305-0801, Japan.}


\begin{abstract}
\noindent
Quantum resonance, {\it i.e.}, amplification in transition probability available under certain conditions, offers a powerful means for determining fundamental quantities in physics, including the time duration of the second adopted in the SI units and neutron's electric dipole moment which is directly linked to CP violation.
We revisit two of the typical examples, the Rabi resonance and the Ramsey resonance, and show that both of these represent the weak value amplification when involving significant enhancement of transition probabilities and that near the resonance points they share exactly the same behavior of transition probabilities except for the measurement strength whose difference leads to the known advantage of the Ramsey resonance in the sensitivity.
Conversely, as a by-product of the relationship, we may measure the weak value through quantum resonance.  In fact, we argue that previous measurements of neutron electric dipole moment based on the Ramsey resonance have potentially determined the weak value of neutron's spin with much higher precision than the conventional weak value measurement.
\end{abstract}

\maketitle

\section{Introduction}
\label{sec:1}

One of the major goals of science is to determine fundamental physical quantities of nature, and to this end the phenomena of resonance are often invoked as a powerful means for measuring the quantities.    
Resonance occurs ubiquitously when a certain condition is met in oscillations, as we are all familiar with acoustic resonance of strings or electrical resonance in tuned circuits.  
As such, its characteristics have been widely exploited for various purposes, including the measurement of frequency of sound waves via acoustic resonance or realization of filters in electrical circuits via electrical resonance.
In quantum physics, resonance is particularly useful to determine physical parameters accurately.  
For instance, in the SI units~\cite{sibrochure,247071} the time dulation of the second is defined  from the ground-state hyperfine transition frequency of the cesium 133 atoms,  
and in particle physics the fundamental parameters such as the neutron electric dipole moment (EDM) in the Standard Model are constrained~\cite{Abel:2020pzs} significantly. 

The quantum resonance emerges as a phenomenon of amplification in transition probability between initial and final states.   
One of the most well-known quantum resonances is the magnetic resonance.  It was first studied by Rabi in 1937 in a general setting~\cite{Rabi:1937dgo} and was subsequently used to measure nuclear magnetic moments~\cite{PhysRev.53.318,PhysRev.53.495,Rabi:1939jho}.
Following this, in 1950 a new mechanism of resonance was devised theoretically by Ramsey~\cite{Ramsey:1950tr} and also confirmed experimentally.  
This Ramsey resonance has an advantage over the Rabi resonance in that it possesses 
a smaller half-width of resonance (0.6 times that of the Rabi resonance) and accordingly higher sensitivity to deviations from its resonance point.
Another advantage of the Ramsey resonance is that the resonance condition is more robust against disturbance of external electric/magnetic fields compared to the Rabi resonance.

As for the accurate (precision) measurement of fundamental parameters, we also have the approach of the weak measurement proposed in 1988 by Aharonov, Albert, and Vaidman~\cite{Aharonov:1988xu} in which one measures a physical quantity weakly on the condition of reaching a prescribed state called the post-selected state at the end of transition.   
The measured value in the procedure is then called the weak value, which offers novel interpretations of physical reality in the time-symmetric description of the measurement process involving the post-selection~\cite{Aharonov2008}. 
With this property, the weak value can play an important role in inferring information about observables without disturbing the system and, for this reason, the weak value has been used to elucidate quantum paradoxes~\cite{aredes2024association}, including the Three Box Paradox~\cite{Aharonov1991CompleteDO}, Danan's paradox on the location of photons~\cite{PhysRevLett.111.240402}, and the quantum Cheshire Cat~\cite{Denkmayr2014,aharonov2021dynamical,aharonov2024angular,Ghoshal:2022bnc}.  

On the more practical side, 
an optical setup was devised to demonstrate that the weak measurement potentially serves as a technique for amplifying the weak value \cite{PhysRevLett.66.1107}.
Subsequently, the weak value amplification was applied to precision measurements such as the observation of the spin Hall effect of light~\cite{hosten2008observation} and the detection of ultrasensitive beam deflection in a Sagnac interferometer~\cite{PhysRevLett.102.173601} (for a review, see {\it e.g.}, Ref.~\cite{Dressel_2014}).
A notable aspect of the weak value amplification is that it is achieved at the cost of suppressing the transition probability between initial and final states.  
In fact, the quantum resonance and the weak value amplification share common characteristics, {\it e.g.}, a transition probability between initial and final states is amplified or suppressed under the conditions of their occurrence.
Although these two approaches of amplification have been extensively studied in the context of precision measurements, the relation between them has remained unnoticed to this day.
This prompts us to revisit the quantum resonance in light of the weak measurement and see the possible connection between the two.

In this paper, we revisit both the Rabi and the Ramsey resonances and point out that the two resonances indeed represent the weak value amplification when the transition probabilities are significantly enhanced and that around their resonance points the behaviors of transition probabilities are exactly the same except for the sensitivity which is proportional to the measurement strength.
In more detail, we show that the Ramsey resonance can amount to the weak measurement $\pi/2\simeq 1/0.6$ times stronger than the Rabi resonance at most, which is consistent with the relation of the half-width mentioned above.  
As a by-product, this allows us to measure the imaginary component of the weak value through the Ramsey resonance. 
It is argued that previous experiments of neutron EDM have potentially determined the imaginary component of the weak value of the spin of neutrons with higher precision than the standard weak value measurement via neutron beams by three orders of magnitude.

This paper is organized as follows:
In Sec.~\ref{sec:2}, we first present the basic idea on the relation between quantum resonance and the weak measurement and then provide necessary formulae of the weak measurement of two-level systems involving a post-selection, which will be used in the following sections when we evaluate the resonances in light of the weak measurement.
In Sec.~\ref{sec:3}, we briefly review the Rabi and Ramsey resonances and then present a unified interpretation of these resonances in terms of the weak measurement. 
In Sec.~\ref{sec:4}, we propose a possible method of measuring neutron EDM through the weak value using the Ramsey resonance technique.
Sec.~\ref{sec:diss} is then devoted to discussions of feasibility of our method in comparison with the weak measurement implemented earlier.
We finish with our summary and outlooks in Sec.~\ref{sec:5}.

\section{Preliminaries}
\label{sec:2}
%
\subsection{Basic idea}
The basic idea relating quantum resonance with the weak measurement is rather simple.  
To see this, we first recall how resonance is treated in 
time-dependent perturbation theory in quantum mechanics (see {\it e.g.,} Refs.~\cite{schiff1955quantum,Griffiths_Schroeter_2018}).  We start with a time-dependent Schr\"{o}dinger equation,
\begin{align}
    i \frac{d}{dt}{|\psi(t)\rangle}=H(t) {|\psi(t)\rangle}, \quad H(t) = h_0+v(t),
    \label{eq:peSch}
\end{align}
where $h_0$ is independent of time while $v(t)$ contains a term representing a time-dependent perturbative effect.  We put $\hbar=1$ throughout this paper for brevity.
Using the expansion ${|\psi(t)\rangle} = \sum_n a_n(t)e^{-i E_n t}{|n\rangle}$ in terms of eigenstates $h_0 {|n\rangle}=E_n{|n\rangle}$ of $h_0$, we rewrite Eq.~\eqref{eq:peSch} as
\begin{align}
    \frac{da_k}{dt}=-i \sum_n {\langle k|v(t)|n\rangle}a_n e^{i\omega_{kn}},\label{eq:akeq}
\end{align}
where $\omega_{kn}= E_k-E_n$.
For the initial condition $a_k(0)=\delta_{km}$, Eq.~\eqref{eq:akeq} is solved, up to the first order of $v(t)$, as
\begin{align}
    a_k(t)\simeq \delta_{km}-i \int_{0}^{t}{\langle k|v(t')|m\rangle} e^{i \omega_{km}t'}dt'.\label{eq:sol1}
\end{align}

Now, let us assume that the term $v(t)$ depends harmonically on time such that ${\langle k|v(t) |k'\rangle} = 0$ for all $k, k'$ except for a particular $k \ne m$ for which 
\begin{align}
{\langle k|v(t) |m\rangle}={\langle m|v(t) |k\rangle} = 2 \omega_1\cos\omega t,
\end{align}
with constant parameters $\omega_1$ and $\omega$.
Then, to such $k$, from Eq.~\eqref{eq:sol1} we obtain
\begin{align}
    a_k(t)= -\omega_1\left(
    \frac{e^{i (\omega_{km}+\omega)t}-1}{\omega_{km}+\omega}+
    \frac{e^{i (\omega_{km}-\omega)t}-1}{\omega_{km}-\omega}
    \right).\label{eq:akt}
\end{align}
This shows that the amplitude $a_k(t)$ becomes significant only when either one of the two denominators is close to zero.
Assuming no energy degeneracy, and focusing on the parameter regions of $\omega$ around one of the resonance points, say, $\omega = \omega_{km}$ for some particular $k$, we see that only two states, ${|m\rangle}$ and ${|k\rangle}$, enter into the time evolution of our concern.  This allows us to consider the system composed essentially of the two states ${|m\rangle}$ and ${|k\rangle}$ only.
Indeed, under these circumstances, we have the approximated expressions,
\begin{align}
    a_k(t) \simeq - \omega_1\frac{e^{i (\omega_{km}-\omega)t}-1}{\omega_{km}-\omega}, \quad a_m(t) \simeq 1,
\end{align}
while $a_n(t)\simeq 0$ for $n\neq k,m$.
Correspondingly, we obtain
\begin{align}
    {\langle k|\psi(t)\rangle}&\simeq -\omega_1 \frac{e^{i (\omega_{km}-\omega)t}-1}{\omega_{km}-\omega} e^{-iE_k t},\label{eq:pe1}
    \\
    {\langle m|\psi(t)\rangle}&\simeq e^{-iE_m t},\label{eq:pe2}
\end{align}
and ${\langle n|\psi(t)\rangle}\simeq 0$ for $n\neq k,m$.
The transition probability from ${|m\rangle}$ to ${|k\rangle}$ then turns out to be
\begin{align}
    {\rm Pr}_{m\to k}:=\left|{\langle k|\psi(t)\rangle}\right|^2 \simeq \frac{4 \omega_1^2 \sin^2 \frac{1}{2}\left(\omega_{km}-\omega\right)t}{\left(\omega_{km}-\omega\right)^2}.\label{eq:conres}
\end{align}
This is a standard result exhibitting a peak in the transition probability around the resonance point $\omega=\omega_{km}$, which holds when $\omega_1$ is small and treated perturbatively.

Next, let us reformulate the time evolution \eqref{eq:peSch} in a more explicit form.  This is possible because we are working, in effect, on the transitions involving only the two states, ${|m\rangle}$ and ${|k\rangle}$, where the system is described by a two-dimensional Hilbert space spanned by the two states.  With this in mind, we put ${|m\rangle} = (1, 0)^{t}$  
and ${|k\rangle} = (0, 1)^{t}$ (where \lq $t$\rq\ denotes the transpose) which enables us to write the Hamiltonian in \eqref{eq:peSch} in the form, 
\begin{align}
    H(t)=\bar E-\frac{\omega_{km}}{2}\sigma_3 
    &+\omega_1\left(\cos\omega t \sigma_1-\sin\omega t \sigma_2 \right)\notag
    \\
    &+\omega_1 \left(\cos\omega t \sigma_1+\sin\omega t\sigma_2
    \right),\label{eq:sbas}
\end{align}
with the average energy $\bar E = (E_k+E_m)/2$ and the Pauli matrices $\sigma_i$, $i=1,2,3$.  
As we did above, we consider the parameter regions around the resonance point $\omega=\omega_{km}$ and adopt the rotating wave approximation in which the last term of Eq.~\eqref{eq:sbas} is omitted.  This leads us to 
\begin{align}
     H(t)=\bar E-\frac{\omega_{km}}{2}\sigma_3
    +\omega_1\left(\cos\omega t \sigma_1-\sin\omega t \sigma_2 \right).
    \label{eq:hamzero}
\end{align}
The Hamiltonian describing the Rabi resonance (\ref{eq:hamRabi}) and partly that of the Ramsey resonance (\ref{eq:HRamsey1}) discussed later take precisely this form \eqref{eq:hamzero} for $\bar E = 0$ and $\omega_{km} = \omega_0$.

It is evident \cite{Griffiths_Schroeter_2018} that this approximation is equivalent to the approximation we have just used when we omit the last term in (\ref{eq:akt}).
In a rotating coordinate frame ${|\psi'(t)\rangle}=e^{-i\omega t\sigma_3/2}{|\psi(t)\rangle}$,
the Schr\"{o}dinger equation (\ref{eq:peSch}) is recast into 
\begin{align}
    i\frac{d}{dt}{|\psi'(t)\rangle}=H'{|\psi'(t)\rangle},
    \label{eq:schIII0}
\end{align}
with the time-independent Hamiltonian,
\begin{align}
    H' = \bar E +\omega_1 \left(\sigma_1+\frac{\omega-\omega_{km}}{2\omega_1}\sigma_3 \right).
    \label{eq:HamRabi0}
\end{align}
The time evolution is then given by
\begin{align}
    {|\psi(t)\rangle}=e^{i\omega t \sigma_3/2}e^{-iH' t}{|\psi(0)\rangle}.
    \label{eq:timeIII0}
\end{align}
For $|\psi(0)\rangle = |m\rangle$ we find
\begin{align}
\!\!\!{\langle k|\psi(t)\rangle}&=e^{-i E_kt}{\langle k|e^{-it \omega_1 \left(\sigma_1 +(\omega-\omega_{km})\sigma_3/2\omega_1\right)}|m\rangle},\label{eq:ge1}
    \\
\!\!\!{\langle m|\psi(t)\rangle}&=e^{-i E_m t}{\langle m|e^{-it \omega_1 \left(\sigma_1 +(\omega-\omega_{km})\sigma_3/2\omega_1\right)}|m\rangle}.\label{eq:ge2}
\end{align}
Note that here we no longer need the perturbative approximation used previously for $\omega_1$.   Indeed,  the amplitudes 
\eqref{eq:ge1} and \eqref{eq:ge2} reduce, respectively, to \eqref{eq:pe1} and \eqref{eq:pe2} in the perturbative regime.   This can be readily confirmed by evaluating the former values to find that they agree with the latters up to the first order of $\omega_1$.
Clearly, the transition probability from ${|m\rangle}$ to ${|k\rangle}$ is found to be 
\begin{align}
    {\rm Pr}_{m\to k}&=\left|{\langle k|e^{-it \omega_1 \left(\sigma_1 +(\omega-\omega_{km})\sigma_3/2\omega_1\right)}|m\rangle}\right|^2
    \nonumber \\
   & =1-{\rm Pr}_{m\to m}.
    \label{eq:Pr2}
\end{align}
This result \eqref{eq:Pr2} shows that the resonance arises at $\omega=\omega_{km}$, and that when one of the two probabilities, ${\rm Pr}_{m\to k}$ and ${\rm Pr}_{m\to m}$, tends to unity then the other necessarily tends to zero.
This (rather trivial) fact that suppression and amplification occurs simultaneously when the resonance condition $\omega=\omega_{km}$ is fulfilled indicates that in quantum resonance the two phenomena are the two sides of the same coin.

To see the connection with the weak measurement, recall that the weak measurement is described by a time evolution of a quantum system involving perturbative effects representing the measurement process under post-selection.
Here, we consider a slight shift $\epsilon$ from the resonance point $\omega=\omega_{km}$ in $H'$ in \eqref{eq:HamRabi0} as  $\epsilon=\omega-\omega_{km}$, and treat it as a parameter of the perturbative effect caused by some external or environmental intervention. 
In view of the time evolution \eqref{eq:timeIII0}, we may consider the effective total Hamiltonian $H$ fulfilling the condition, 
\begin{align}
e^{-iH t} = e^{i\omega t \sigma_3/2}e^{-iH' t},
\label{eq:totalham}
\end{align}
and split it as $H = H_0 + V$ with $V=\epsilon \sigma_3/2$ being the perturbative term proportional to $\epsilon$.  
It is now obvious that this is nothing but the standard situation where the effect of the perturbation given by $V$ can be considered as an interaction that implements a direct weak measurement on the change of transition probability \cite{PhysRevLett.111.023604,PhysRevA.101.042117,RevModPhys.86.307,Mori:2021avk}.  
In fact, given a pair of 
pre-selected state ${|\psi_i\rangle}$ and the post-selected state ${|\psi_f\rangle}$, 
the weak value associated with $V$ is retrieved from the transition amplitude,  
\begin{align}
{\langle \psi_f|}e^{-i H t}{|\psi_i\rangle} = {\langle \psi_f|}e^{-i (H_0 + V) t}{|\psi_i\rangle}.
\label{eq:trpro}
\end{align}
We shall see this shortly below in full generality, but in a particularly simple case where $H_0$ can be ignored, it is immediate to see that, to the first order of perturbation, the amplitude becomes 
\begin{align}
{\langle \psi_f|}e^{-i H t}{|\psi_i\rangle} \simeq  {\langle \psi_f|}{\psi_i\rangle}\left(1 -i \epsilon \sigma_3^W/2\right), \quad
\label{eq:wvap}
\end{align}
where 
\begin{align}
\sigma_3^W :=\frac{\langle \psi_f|\sigma_3|\psi_i\rangle}{\langle \psi_f|\psi_i\rangle},
\end{align}
is the weak value of $\sigma_3$.
Despite the simplicity, many instances follow the above type of time evolution; see, {\it e.g.}, Refs.~\cite{Denkmayr2014,hosten2008observation,PhysRevLett.102.173601,Song:2023uht,zhang2015precision,qiu2017precision}, where one observes from \eqref{eq:wvap} that the imaginary part of the weak value ${\rm Im}\,\sigma_3^W$ affects the magnitude of transition amplitude whereas the real part ${\rm Re}\,\sigma_3^W$ affects its phase.  The combination of these in the form of the modulus then represents the susceptibility of the system under the perturbation $V$.  In fact, 
this is a common feature of the weak value seen even in indirect measurement of weak values~\cite{sponar2015weak} (for a comprehensive review containing the discussion on the direct measurement, see Ref.~\cite{Dressel_2014}). Unfortunately, we do not afford this feature in our scheme of direct weak measurement which is more general and tailored for the quantum resonances discussed later. 

Before going over to the discussion of our scheme, we note that, for the choice $|\psi_i\rangle = |m\rangle$ and $|\psi_f\rangle = |k\rangle$, we have 
$H = H'$ effectively, since the additional factors in \eqref{eq:totalham} yield only a phase which can be ignored in the transition probability.
Another point to be noted is that, upon a particular set of points such as  $\omega_1 t = \pi/2$, 
the transition probability is suppressed via the quantum resonance, which implies that under these circumstances the weak value 
 (which has the factor $\langle \psi_f|\psi_i\rangle$ in its denominator) becomes large near the resonance points.  In other words, 
the weak value amplification takes place around the resonance points when the suppression of the transition probability occurs.  
To find the explicit form of the weak value, we need to take account of the possible \lq rotation\rq\ in the two-dimensional Hilbert space 
due to the composition of the operators involved.  The detail will be described next.

\subsection{General formulae}
For the convenience of later sections, we shall summarize the formulae for time evolution of two-level systems with post-selections, generalizing the situation mentioned above.  For the readers who are familiar with the formulae or more interested in the outcomes of applications may go directly to Sect.~III.  

First, we consider a system described by a time-independent Hamiltonian,
\begin{align}
    H=H_0+V,
\label{eq:Ham1}
\end{align}
where $H_0$ denotes a part dominantly governing the time evolution of the system, and $V$ is a part describing a perturbative effect on the system.  
We also assume that $V$ is an operator representing a direct measurement process in the system.
In contrast to indirect measurements, direct measurements require no system of measurement device and extract information through the change  in the transition probability between $V=0$ and $V\neq0$. 
As explained later with some examples, the measurement operator $V$ is defined appropriately according to the situations we have. 
We parametrize the two terms of the Hamiltonian \eqref{eq:Ham1} in the usual manner as
\begin{align}
    H_0 &= h \left(n_0^{(h)}+ \vec{n}^{(h)}\cdot \vec{\sigma}\right),~~~V = v\left(n_0^{(v)}+ \vec{n}^{(v)}\cdot \vec{\sigma}\right),
    \label{eq:h0V}
\end{align}
where $h, v$ are real positive values, 
$n_0^{(h)}, n_0^{(v)}$ stand for the part proportional to the 2 $\times$ 2 identity matrix, and  
$\vec{n}^{(h)}\cdot \vec{\sigma}=\sum_{i=1}^3 n_{i}^{(h)} \sigma_i$ and $\vec{n}^{(v)}\cdot \vec{\sigma}=\sum_{i=1}^3 n_i^{(v)} \sigma_i$.  
The vectors $\vec{n}^{(h)}=(n_1^{(h)},n_2^{(h)},n_3^{(h)})$ and $\vec{n}^{(v)}=(n_1^{(v)},n_2^{(v)},n_3^{(v)})$ 
are normalized as $\vec{n}^{(h)}\cdot \vec{n}^{(h)\ast}=1$ and $\vec{n}^{(v)}\cdot \vec{n}^{(v)\ast}=1$.  
These components $n_0^{(h)}, n_0^{(v)}$ and $n_i^{(h)}, n_i^{(v)}$ for $i=1,2,3$ may take complex values when non-hermitian time evolution is considered in a situation where, for instance, an absorber is introduced in the system.
The reason why we also consider the non-hermitian time evolution here is that, in our scheme of weak measurement designed for quantum resonances, 
the non-hermitian time evolution is required for the real part of the weak values to arise, as 
will be seen explicitly below.

Let ${|\psi(t)\rangle}$ be a quantum state of the system at time $t$.
The time evolution of the state between $t_0$ and $t_0+t$ is given by
\begin{align}
{|\psi(t_0+t)\rangle}&=e^{-i Ht}{|\psi_i\rangle},\label{eq:time1}
\end{align}
where ${|\psi_i\rangle}:={|\psi(t_0)\rangle}$.
%
%
%
%
%
%
Using the standard formula, 
\begin{equation}
e^{i\vec{x}\cdot \vec{\sigma}} = \cos \sqrt{(\vec{x})^2} +\frac{i\vec{x}\cdot \vec{\sigma}}{\sqrt{(\vec{x})^2}} \sin \sqrt{(\vec{x})^2},
\end{equation}
it is straightforward to show that the time evolution can be approximated, up to the first order of $v$, as
\begin{widetext}
\begin{align}
    {|\psi(t_0+t)\rangle}&\simeq e^{-i(h n_0^{(h)}+vn_0^{(v)})t}\left[
    e^{-ih\sigma_{h} t}\left(
    1-ivt \frac{\vec{n}^{(h)}\cdot \vec{n}^{(v)}}{(\vec{n}^{(h)})^2}\sigma_{h}
    \right)-i\frac{v}{h}\frac{1}{\sqrt{(\vec{n}^{(h)})^2}}\left(\sigma_{v}-\frac{\vec{n}^{(h)}\cdot \vec{n}^{(v)}}{(\vec{n}^{(h)})^2}\sigma_{h}\right)\sin \left(h \sqrt{(\vec{n}^{(h)})^2}t\right)
    \right]{|\psi_i\rangle}
    \nonumber 
    \\
    &= e^{-ivn_0^{(v)}t}\left[
    e^{-i H_0 t}
    \left(1-ivt \frac{\vec{n}^{(h)}\cdot \vec{n}^{(v)}}{(\vec{n}^{(h)})^2}\sigma_{h} \right)
    -i \frac{v}{2h}[\sigma_{a},e^{-iH_0 t}]
    \right]{|\psi_i\rangle},
    \label{eq:time2}
\end{align}
\end{widetext}
where we have introduced the shorthands, 
\begin{equation}
\sigma_{h}:=\vec{n}^{(h)}\cdot \vec{\sigma}, \quad \sigma_{v}:=\vec{n}^{(v)}\cdot \vec{\sigma}, \quad \sigma_{a}:=\vec{n}^{(a)}\cdot \vec{\sigma},
\end{equation}
with $\vec{n}^{(a)}=(n_1^{(a)},n_2^{(a)},n_3^{(a)})$ satisfying the normalization condition $\vec{n}^{(a)}\cdot \vec{n}^{(a)\ast}=1$.  The last one $\sigma_{a}$ is chosen in such a way that it fulfills the relation with the former two as 
\begin{align}
[\sigma_{a}, \sigma_{h}]=2i
\left(\sigma_{v}-\frac{\vec{n}^{(h)}\cdot \vec{n}^{(v)}}{(\vec{n}^{(h)})^2}\sigma_{h}\right).
\label{eq:condsa}
\end{align}
In what follows we focus on the case where $v$ takes a small value so that our approximation up to the first order of $v$ is valid.
Note that the second terms in the first and the second lines of Eq.~\eqref{eq:time2} vanish for $\vec{n}^{(h)}\propto \vec{n}^{(v)}$ for which we have  $\sigma_{v}=(\vec{n}^{(h)}\cdot \vec{n}^{(v)})/(\vec{n}^{(h)})^2\sigma_{h}$ and $[\sigma_a,e^{-iH_0 t}]=0$.
After the time evolution, we perform the post-selection, that is, we restrict ourselves to the case where the state at the end is found to be the post-selected state ${|\psi_f\rangle}$ we have specified in advance.
This measurement procedure on the perturbative effect through the transition probability involving the post-selection (with the approximation up to the first order of $v$) is referred to as {\it the (direct) weak measurement}.
The transition amplitude from ${|\psi_i\rangle}$ to ${|\psi_f\rangle}$ then reads
\begin{widetext}
\begin{align}
    {\langle \psi_f|\psi(t_0+t)\rangle}= {\langle \psi_f|}e^{-i H t}{|\psi_i\rangle}
\simeq e^{-ivn_0^{(v)}t}{\langle \psi_f |e^{-i H_0 t}|\psi_i\rangle}
    \left(
    1-i vt \frac{\vec{n}^{(h)}\cdot \vec{n}^{(v)}}{(\vec{n}^{(h)})^2}\sigma^W_{h}
-i\frac{v}{2h}\left(\sigma^W_{a,L}-\sigma^W_{a,R}\right)
    \right)
    ,\label{eq:amp1}
\end{align}
where 
\begin{align}
    \sigma_{h}^W&:=\frac{{\langle \psi_f|}\sigma_{h} e^{-iH_0 t}{|\psi_i\rangle}}{{\langle \psi_f|}e^{-iH_0 t}{|\psi_i\rangle}}= \frac{{\langle \psi_f|}e^{-iH_0t}\sigma_{h} {|\psi_i\rangle}}{{\langle \psi_f|}e^{-iH_0t}{|\psi_i\rangle}},~~~    \sigma^W_{{a},L}:=\frac{{\langle \psi_f|\sigma_{a}e^{-iH_0t}|\psi_i\rangle}}{{\langle \psi_f|e^{-iH_0t}|\psi_i\rangle}},~~~\sigma^W_{{a},R}:=\frac{{\langle \psi_f|e^{-iH_0t}\sigma_{a}|\psi_i\rangle}}{{\langle \psi_f|e^{-iH_0t}|\psi_i\rangle}}.\label{eq:weakI}
\end{align}
are regarded as {\it the weak values} corresponding to $\sigma_{h}$ and $\sigma_{a}$, respectively.  For the latter the weak value may take different values at the initial and final times of the transition and hence requires the distinction with the labels \lq R\rq\ and \lq L\rq, while for the former the two coincide.
With these, the transition probability from ${|\psi_i\rangle}$ to ${|\psi_f\rangle}$ is writtien as
\begin{align}
    {\rm Pr}_{i\to f}\left(v\right)&:=\left|{\langle \psi_f|\psi(t_0+t)\rangle}\right|^2\notag
    \\
    &\simeq {\rm Pr}_{i\to f}\left(0\right)e^{2vt\,{\rm Im}\,n_0^{(v)}}
    \left[
    1+2vt \left(
    {\rm Im}\,\left(\frac{\vec{n}^{(h)}\cdot \vec{n}^{(v)}}{(\vec{n}^{(h)})^2}\right){\rm Re}\,\sigma^W_{h}+{\rm Re}\,\left(\frac{\vec{n}^{(h)}\cdot \vec{n}^{(v)}}{(\vec{n}^{(h)})^2}\right){\rm Im}\,\sigma^W_{h}
    \right)
    +\frac{v}{h} \left({\rm Im}\,\sigma^W_{{a},L}-{\rm Im}\,\sigma^W_{{a},R}\right)
    \right],\label{eq:pr1}
\end{align}
\end{widetext}
with ${\rm Pr}_{i\to f}(0)=\left|{\langle \psi_f |e^{-iH_0 t}|\psi_i\rangle}\right|^2$.
Eq.~\eqref{eq:pr1} shows that the non-hermitian time evolution is needed to obtain the real part of the weak value, while the usual hermitian time evolution is enough to obtain the imaginary part.

At this point, it is perhaps worthwhile to mention the difference between the direct weak measurement used in this paper and the more conventional indirect weak measurement considered in most of the previous works.  Namely, 
in the indirect weak measurement, we introduce, in addition to the target system, a probe system with which the physical value of the target system is measured.  In the context of weak measurement, the weak value of the target system is obtained by observing a change in the position of the pointer, for instance, of the probe system by some strong (ideal) measurement.  Obviously, the retrieval of the weak value depends heavily on the way the physical operator of the target system couples to the pointer operator in the probe system.  In contrast, such complication does not arise in the direct measurement where we only uses the target system. Yet, even in the direct measurement one has the freedom of choice in the perturbative part of the Hamiltonian, which gives rise to a different meaning to the weak value depending on the choice, as we shall mention shortly.

Below, paying particular attention to extracting the imaginary parts of the weak values, we shall analyze the time evolution in more detail for the typical two cases: $\vec{n}^{(h)}\times \vec{n}^{(v)}=0$ and $\vec{n}^{(h)}\cdot \vec{n}^{(v)}=0$ both with $n_0^{(v)}=0$.
%

\subsubsection{Commutative case: $\vec{n}^{(h)}\times \vec{n}^{(v)}=0$}
%
In the case of $\vec{n}^{(h)}\times \vec{n}^{(v)}=0$, {\it i.e.}, $\vec{n}^{(h)}\propto\vec{n}^{(v)}$, the two operators $H_0$ and $V$ commute  to each other, $[H_0, V]=0$.
As will be explained later, the Ramsey resonance corresponds to this case.
For $n_0^{(v)}=0$, up to the first order of $v$, the time evolution of Eq.~\eqref{eq:time2} becomes
\begin{align}
    {|\psi(t_0+t)\rangle}
    &\simeq e^{-i H_0 t} \left(1-iv t \sigma_v\right){|\psi_i\rangle},
    \label{eq:timeI1}
\end{align}
%
%
and hence
\begin{align}
    {\langle \psi_f|\psi(t_0+t)\rangle}\simeq {\langle \psi_f|}e^{-iH_0t}{|\psi_i\rangle} \left(1-iv t \sigma_{v}^W\right)\label{eq:ampI}.
\end{align}
%
%
%
Substituting Eq.~\eqref{eq:ampI} into Eq.~\eqref{eq:pr1}, we end up with
\begin{align}
    {\rm Pr}_{i\to f}\left(v\right)
    &\simeq {\rm Pr}_{i\to f}\left(0\right) \left(1+2vt\,{\rm Im}\, \sigma_{v}^W\right).\label{eq:PrI}
\end{align}
This represents the effects of the measurement involving the post-selection up to the first order of $v$, {\it i.e.,} the weak measurement on the transition probability. 
From Eq.~\eqref{eq:PrI}, the imaginary part of the weak value of $\sigma_v$ is given by
\begin{align}
    {\rm Im}\,\sigma_{v}^W =
    \frac{1}{2{\rm Pr}_{i\to f}\left(0\right)}\frac{d{\rm Pr}_{i\to f}\left(v\right)}{d(vt)}\bigg|_{v=0}.\label{eq:weakI2}
\end{align}
This formula shows that the weak value ${\rm Im}\,\sigma_{v}^W$ can be measured by combining observable quantities ${\rm Pr}_{i\to f}\left(0\right)$ and ${\rm Pr}_{i\to f}\left(v\right)$.
The parameter $vt$ appearing in Eq.~\eqref{eq:weakI2} is recognized as the {\it measurement strength}, and our calculations carried out to the first order of $vt$ are deemed to be applicable for the weak measurements satisfying $vt \ll1$.

The result \eqref{eq:weakI2} also shows that the imaginary part of the weak value ${\rm Im}\,\sigma_{v}^W$ signifies the susceptibility with respect to the interaction $V$ in the transition amplitude. 
This is contrasted to the simple case discussed earlier for \eqref{eq:wvap}, where we have observed that the imaginary part of the weak value represents the susceptibility in the amplitude while the real part represents it in the phase under the interaction $V$.  The fact that different meanings can be associated to (the real and imaginary parts of) the weak value is no surprise, given that the appearance of the weak value depends on the choice of the terms $H_0$ and $V$ in the Hamiltonian $H$ as well as on the choice of the perturbative parameter we introduce.  In other words, there is no definite characterization of the physical meaning of the weak value, which is determined according to the scheme of weak measurement one employs.  The physical significance of the weak value can be found only with respect to the perturbative parameter and the associated operator which controls the weak measurement in the scheme.
%

\subsubsection{Non-commutative case: $\vec{n}^{(h)}\cdot \vec{n}^{(v)}=0$}
\label{sec:non}
%
In the case of $\vec{n}^{(h)}\cdot \vec{n}^{(v)}=0$, the two operators $H_0$ and $V$ are not commutative, $[H_0,V]\neq 0$.
As we shall see later, the Rabi resonance falls in this case.
For $n^{(v)}_0=0$, up to the first order of $v$, the time evolution of Eq.~\eqref{eq:time2} is expressed as
\begin{align}
    {|\psi(t_0+t)\rangle}\simeq \left(e^{-i H_0t}-i\frac{v}{2h} [\sigma_{a},e^{-iH_0t}]\right){|\psi_i\rangle},
    \label{eq:timeI}
\end{align}
%
%
which implies
\begin{align}
    {\langle \psi_f|\psi(t_0+t)\rangle}\simeq {\langle \psi_f|e^{-iH_0t}|\psi_i\rangle}
    \left(1-i\frac{v}{2h}\left( \sigma_{{a},L}^W- \sigma_{a,R}^W \right)\right).
    \label{eq:ampII}
\end{align}
Substituting Eq.~\eqref{eq:ampII} into Eq.~\eqref{eq:pr1}, we find
\begin{align}
    {\rm Pr}_{i\to f}\left(v\right)
    &\simeq {\rm Pr}_{i\to f}\left(0\right) 
    \left(1+\frac{v}{h} \left({\rm Im}\,\sigma^W_{{a},L}-{\rm Im}\,\sigma^W_{{a},R}\right)\right).\label{eq:PrII}
\end{align}
As before, Eq.~\eqref{eq:PrII} is rewritten as
\begin{align}
    {\rm Im}\,\sigma^W_{a,L}-{\rm Im}\,\sigma^W_{a,R}=
    \frac{1}{{\rm Pr}_{i\to f}\left(0\right)}\frac{d{\rm Pr}_{i\to f}\left(v\right)}{d(v/h)}\bigg|_{v=0}.\label{eq:weakII}
\end{align}
From Eq.~\eqref{eq:weakII}, the parameter $v/h$ is now recognized as the measurement strength of the weak measurement which provides the difference ${\rm Im}\,\sigma^W_{a,L}-{\rm Im}\,\sigma^W_{a,R}$.
The above results are applicable for the weak measurement satisfying the condition $v/h \ll1$.

Comparing Eq.~\eqref{eq:PrII} with Eq.~\eqref{eq:PrI}, it is clear that the transition probabilities between the initial and the final states are determined by distinct weak values depending on the commutative and non-commutative cases.
Despite the difference,  we shall observe later in Section.~\ref{sec:3} that there is a close parallel in  
the transition probabilities between the Ramsey resonance and the Rabi resonance, which correspond to the  
commutative and the non-commutative case of the weak measurement, respectively.
\section{Resonance}
\label{sec:3}
%
%

Quantum resonance is a phenomenon in which the transition amplitude is amplified when a specific condition holds.
Here, using the formulae provided in Sec.~\ref{sec:2}, we revisit the Rabi and Ramsey resonances to see that they admit a unified interpretation in light of the weak measurements associated with the two resonances.  

\subsection{Rabi resonance}
\label{sec:Rabi}
Consider a system described by the following time-dependent Hamiltonian:
\begin{align}
H_{\rm Rabi}(t):=-\frac{\omega_0}{2}\sigma_3+\omega_1 \left(\cos\omega t\, \sigma_1-\sin\omega t\, \sigma_2\right),
\label{eq:hamRabi}
\end{align}
where $\omega_0, \omega_1$ and $\omega$ are real constants.
This describes a particle spin system under an external magnetic field with $\omega_0$ being the product of a statistic magnetic field along the $z$ axis and the particle's magnetic moment.  In addition, the spin is under the influence of a rotating magnetic field with an angular frequency $\omega$ on the $x-y$ plane, and $\omega_1$ represents the of the magnitude of the product of the rotating field and the particle's magnetic moment.
The Rabi resonance occurs in this setting and is commonly referred to as magnetic resonance.

The time evolution of a state ${|\psi(t)\rangle}$ is described by the Schr\"{o}dinger equation,
\begin{align}
    i\frac{d}{dt}{|\psi(t)\rangle}=H_{\rm Rabi}(t){|\psi(t)\rangle}.\label{eq:schIII}
\end{align}
In a rotating coordinate frame ${|\psi'(t)\rangle}:=e^{-i\omega t\sigma_3/2}{|\psi(t)\rangle}$, Eq.~\eqref{eq:schIII} is recast into the Schr\"{o}dinger equation,
$
    i{d}{|\psi'(t)\rangle}/{dt}=H'_{\rm Rabi}{|\psi'(t)\rangle},
$
with the time-independent Hamiltonian,
\begin{align}
    H'_{\rm Rabi}:=\omega_1 \left(\sigma_1+\frac{\omega-\omega_0}{2\omega_1}\sigma_3 \right).\label{eq:HamRabi}
\end{align}
The time evolution of the state between $t_0$ and $t_0+t$ is given by
\begin{align}
    {|\psi(t_0+t)\rangle}=e^{i\omega (t_0+t)\sigma_3/2}e^{-iH'_{\rm Rabi} t}e^{-i\omega t_0\sigma_3/2}{|\psi(t_0)\rangle}.\label{eq:timeIII}
\end{align}
We may thus regard $H'_{\rm Rabi}$ as the Hamiltonian $H$ of Eq.~\eqref{eq:Ham1}.

Now we choose our initial state as ${|\psi(t_0)\rangle}={|\pm\rangle}$, which is an eigenstate of $\sigma_3$ satisfying $\sigma_3 {|\pm\rangle} =\pm {|\pm\rangle}$.
Then, from Eq.~\eqref{eq:timeIII}, the transition amplitude from ${|\pm\rangle}$ to ${|\pm\rangle}$ is obtained as
\begin{align}
{\langle \pm| \psi(t_0+t)\rangle}=e^{\pm i\omega t/2} {\langle \pm| e^{-iH'_{\rm Rabi}t}|\pm\rangle}.\label{eq:ampIII}
\end{align}
From Eq.~\eqref{eq:ampIII}, the corresponding transition probability is given by
\begin{align}
    {\rm Pr}_{\pm\to\pm}^{\rm Rabi}=\left|{\langle \pm| e^{-iH'_{\rm Rabi}t}|\pm\rangle}\right|^2.\label{eq:proIII}
\end{align}
A resonance arises when $\omega=\omega_0$ yielding $H'_{\rm Rabi}=\omega_1\sigma_1$, for which we obtain
\begin{align}
    {\rm Pr}_{\pm\to\pm}^{\rm Rabi}=\left|{\langle \pm| e^{-i\omega_1t\sigma_1}|\pm\rangle}\right|^2~{\rm for}~\omega=\omega_0.\label{eq:proIII2}
\end{align}
Then, for $\omega_1t=\pi/2$ the amplitude Eq.~\eqref{eq:ampIII} is suppressed and the probability of Eq.~\eqref{eq:proIII2} reduces to zero.
On the other hand, under the same condition the transition probability from ${|\pm\rangle}$ to ${|\mp\rangle}$ is amplified and becomes one on account of ${\rm Pr}_{\pm\to\mp}^{\rm Rabi}=1-{\rm Pr}_{\pm\to\pm}^{\rm Rabi}$ as mentioned in (\ref{eq:Pr2}).
%
%
Note here that away from the resonance point, the unitary operator $e^{-iH'_{\rm Rabi}t}$ implements a rotation of state around an axis slightly shifted from the z-axis, and the probability of Eq.~\eqref{eq:proIII} cannot be zero even if we tune $\omega_1$. 
From these, $\omega_1t=\pi/2$ represents the condition that the peak of the transition probability ${\rm Pr}_{\pm\to\mp}^{\rm Rabi}$ is maximally enhanced at the resonance point $\omega=\omega_0$.
Deviations from $\omega_1t=\pi/2$ suppress its peak at the resonance point.
In what follows, we mainly consider cases $1/2\ll{\rm Pr}_{\pm\to\mp}^{\rm Rabi}$ in which the peak is significantly enhanced compared with another transition probability ${\rm Pr}_{\pm\to\pm}^{\rm Rabi}=1-{\rm Pr}_{\pm\to\mp}^{\rm Rabi}$ as the resonance phenomena since the resonance phenomenon has its value in the significant amplification of probability.

A key aspect of this resonance is that the probability ${\rm Pr}_{\pm\to \mp}^{\rm Rabi}$ is amplified significantly at $\omega=\omega_0$, {\it e.g.,} when $\omega_1t$ is fixed at $\pi/2$.
This enables us to measure the value of $\omega_0$ through the resonance accurately.
To see how this is achieved, let us imagine an experiment where we have three adjustable parameters $\omega, \omega_1$ and $t$, and one unknown parameter $\omega_0$.
It is clear that one can measure the unknown parameter $\omega_0$ by tuning $\omega$ while keeping $\omega_1$ and $t$ and observing where the peak in the probability occurs, which leads to the value $\omega=\omega_0$ with accuracy determined from the width of the peak.  
This type of measurement process can be regarded as a weak measurement because it extracts information about the slight disturbance of the system around the resonance condition.

We now recosider the above resonance phenomena in the context of the weak measurement.
For our convenience, let us decompose the parameter $\omega_0$ as
\begin{align}
    \omega_0:=\overline{\omega}_0+\epsilon,
    \label{eq:ep1}
\end{align}
where the parameter $\epsilon$ represents a small disturbance to be measured.  We then 
consider a weak measurement of $\epsilon$ via the Rabi resonance in the following setup:  
\begin{itembox}[l]{Weak measurement via Rabi resonance}
\begin{itemize}
    \item Known parameters: $\overline{\omega}_0$, $\omega$, $\omega_1$, and $t$
    \item Unknown parameter: $\epsilon$
    \item Condition of weak measurement: $\epsilon/\omega_1\ll1$
\end{itemize}
\end{itembox}
%
%
%
%
Since the parameter $\epsilon$ characterizes a disturbance of the system with reference to the null case $\epsilon=0$, we here consider $\epsilon/\omega_1\ll1$ to be the condition of the weak measurement. 
With this in mind, we rewrite the Hamiltonian of Eq.~\eqref{eq:HamRabi} as
\begin{align}
    H'_{\rm Rabi}=\omega_1 \sigma_1+\omega_1\left(\phi_{\rm Rabi}+\delta\right)\sigma_3,
    \label{eq:Rabipri}
\end{align}
with
\begin{align}
\phi_{\rm Rabi}:=\frac{\omega-\overline{\omega}_0}{2\omega_1}
   \label{eq:Rabiphase}
\end{align}
and $\delta:= -\epsilon/2\omega_1$.  
Regarding $H'_{\rm Rabi} = H$ in Eq.~\eqref{eq:Ham1}, we choose the two operators $H_0$ and $V$ as
\begin{align}
    H_0&=\omega_1\left( \sigma_1+\phi_{\rm Rabi}\, \sigma_3\right),~~~V=\omega_1\delta\,\sigma_3.\label{eq:HamRab}
\end{align}

In the following, we restrict ourselves to a parameter region near the resonance point where $\phi_{\rm Rabi} \ll 1$ and $\delta\ll 1$ hold.
Up to the first order of $\phi_{\rm Rabi}$ and $\delta$, the coefficients of the Pauli matrices of Eq.~\eqref{eq:h0V} are found to be
\begin{align}
    &h=\omega_1,~~n_0^{(h)}=0,~~\vec{n}^{(h)}= \left(1,0,\phi_{\rm Rabi}\right),\notag
    \\
    &v=\omega_1\delta,~~n_0^{(v)}=0,~~\vec{n}^{(v)}=\left(0,0,1\right).
\end{align}
Since $\sigma_h = \sigma_1+\phi_{\rm Rabi}\, \sigma_3$ and $\sigma_v = \sigma_3$ in our approximation,  
taking $\sigma_a=-\sigma_2$ we find 
\begin{align}
[\sigma_a, \sigma_h]=2i (\sigma_3 + \phi_{\rm Rabi}\, \sigma_1),
\end{align}
which is precisely the condition \eqref{eq:condsa} required for $\sigma_a$ within the same approximation.  
From Eq.~\eqref{eq:time2}, the time evolution of the system is found to be
\begin{widetext}
\begin{align}
    {|\psi(t_0+t)\rangle}&\simeq  \left(
    e^{-iH_0t}+i\frac{\delta}{2} \left[\sigma_2,
    e^{-iH_0t}\right]\right){|\psi_i\rangle}\notag
    \\
    &\simeq  \left(e^{i\phi_{\rm Rabi}\, \sigma_2/2}
    e^{-i\omega_1t\sigma_1}e^{-i\phi_{\rm Rabi}\, \sigma_2/2}+i\frac{\delta}{2} \left[\sigma_2,e^{i\phi_{\rm Rabi}\, \sigma_2/2}
    e^{-i\omega_1t\sigma_1}e^{-i\phi_{\rm Rabi}\, \sigma_2/2}\right]\right){|\psi_i\rangle},\label{eq:RabiTIM}
\end{align}
where we have used $e^{i\phi_{\rm Rabi}\,\sigma_2/2}\sigma_1 e^{-i\phi_{\rm Rabi}\,\sigma_2/2}=\sigma_1\cos\phi_{\rm Rabi}+\sigma_3\sin\phi_{\rm Rabi}$.
Substituting Eq.~\eqref{eq:RabiTIM} into Eq.~\eqref{eq:amp1}, we obtain
\begin{align}
    {\langle \psi_f|\psi(t_0+t)\rangle}\simeq {\langle \psi_f(\phi_{\rm Rabi}) |
    e^{-i\omega_1 t \sigma_1}|\psi_i(\phi_{\rm Rabi})\rangle}\left(1+i\frac{\delta}{2}\left(\sigma^W_{2,L}\left(\phi_{\rm Rabi}\right)-\sigma^W_{2,R}\left(\phi_{\rm Rabi}\right)\right)\right),\label{eq:RABAMP}
\end{align}
where ${|\psi_{i}(\phi_{\rm Rabi})\rangle}:=e^{-i \phi_{\rm Rabi} \sigma_2/2}{|\psi_i\rangle}$, ${|\psi_{f}(\phi_{\rm Rabi})\rangle}:=e^{-i \phi_{\rm Rabi} \sigma_2/2}{|\psi_f\rangle}$, and the weak values are given by
\begin{align}
\sigma^W_{2,L}\left(\phi_{\rm Rabi},\omega_1t\right)&:=\frac{{\langle \psi_f(\phi_{\rm Rabi})|\sigma_2 e^{-i\omega_1t \sigma_1}|\psi_i(\phi_{\rm Rabi})\rangle}}{{\langle \psi_f(\phi_{\rm Rabi})|e^{-i\omega_1t \sigma_1}|\psi_i(\phi_{\rm Rabi})\rangle}},~~~\sigma^W_{2,R}\left(\phi_{\rm Rabi},\omega_1t\right):=\frac{{\langle \psi_f(\phi_{\rm Rabi})| e^{-i\omega_1t \sigma_1}\sigma_2|\psi_i(\phi_{\rm Rabi})\rangle}}{{\langle \psi_f(\phi_{\rm Rabi})| e^{-i\omega_1t \sigma_1}|\psi_i(\phi_{\rm Rabi})\rangle}}.\label{eq:weakR}
\end{align}
Plugging ${|\psi_i\rangle}={|\pm\rangle}$, ${|\psi_f\rangle}={|\pm\rangle}$ and Eq.~\eqref{eq:RABAMP} into Eq.~\eqref{eq:pr1}, we obtain the transition probability,
\begin{align}
    {\rm Pr}_{\pm\to \pm}^{\rm Rabi}(\epsilon)\simeq {\rm Pr}_{\pm\to\pm}^{\rm Rabi}(0)  \left(1+\frac{\epsilon}{2\omega_1}\left({\rm Im}\,\sigma_{2,L}^W\left(\phi_{\rm Rabi},\omega_1t\right)-{\rm Im}\,\sigma_{2,R}^W\left(\phi_{\rm Rabi},\omega_1t\right)\right)\right).
   \label{eq:propm}
\end{align}
\end{widetext}
In particular, for $\omega_1t=\pi/2$, the weak values become
\begin{align}
    \sigma^W_{2,L}\left(\phi_{\rm Rabi},\pi/2\right)=-\sigma^W_{2,R}\left(\phi_{\rm Rabi},\pi/2\right)=-i\cot\phi_{\rm Rabi}.\label{eq:weaksRab}
\end{align}
The imaginary parts of the weak values are generally given as,
\begin{align}
    &{\rm Im}\, \sigma^W_{2,L}\left(\phi_{\rm Rabi},\omega_1 t\right)=-{\rm Im}\,\sigma^W_{2,R}\left(\phi_{\rm Rabi},\omega_1 t\right)\notag
    \\
    &\quad\quad\quad= -\cot \phi_{\rm Rabi}\times \frac{\sin^2 \phi_{\rm Rabi} \tan^2 (\omega_1 t)}{1+\sin^2 \phi_{\rm Rabi} \tan^2 (\omega_1 t)}. \label{eq:Rabi_weak_im}
\end{align}
Also, the probability for $\epsilon =0$ is obtained as
\begin{align}
    {\rm Pr}_{\pm \to\pm}^{\rm Rabi}(0)&=\left|{\langle \psi_f(\phi_{\rm Rabi})|e^{-i\omega_1t \sigma_1}|\psi_i(\phi_{\rm Rabi})\rangle}\right|^2\notag
    \\
    &=\sin^2 \phi_{\rm Rabi}\left(1+\cos^2(\omega_1 t)\cot^2\phi_{\rm Rabi}\right).\label{eq:Rabi_zero}
\end{align}
Note here that Eq.~\eqref{eq:Rabi_zero} hold even for $\epsilon\neq 0$ by replacing $\phi_{\rm Rabi}\to \phi_{\rm Rabi}-\epsilon/2\omega_1$ because the shift from the resonance point $\epsilon$ can be absorbed to $\phi_{\rm Rabi}$.
The transition probability \eqref{eq:propm} then turns out to be
\begin{align}
    {\rm Pr}_{\pm\to\pm}^{\rm Rabi}(\epsilon)\simeq {\rm Pr}_{\pm\to\pm}^{\rm Rabi}(0)  \left(1+\delta_{\rm Rabi}\, {\rm Im}\,\sigma_{2,L}^W\left(\phi_{\rm Rabi},\omega_1 t\right)\right),\label{eq:Prr}
\end{align}
where $\delta_{\rm Rabi}:=\epsilon /\omega_1$ is the measurement strength of the Rabi resonance.
%
%
%
%
%
In the limit $\phi_{\rm Rabi}=0$, the probability \eqref{eq:Prr} vanishes, ${\rm Pr}_{\pm\to \pm}^{\rm Rabi}(\epsilon)=0$, signaling the appearance of resonance~\footnote{The resonance condition remains unchanged up to the first order of $\epsilon$, although it is modified from the second order of $\epsilon$.}.
In this limit, in particular, under the condition $\tan^2 (\omega_1t)\geq |\phi_{\rm Rabi}|^{-1}$, the weak values \eqref{eq:Rabi_weak_im} are simultaneously amplified, {\it i.e.,} $|{\rm Im}\, \sigma^W_{2,L}\left(\phi_{\rm Rabi},\omega_1 t\right)|=|{\rm Im}\, \sigma^W_{2,R}\left(\phi_{\rm Rabi},\omega_1 t\right)|\geq 1$.
The condition $\tan^2 (\omega_1t)\geq |\phi_{\rm Rabi}|^{-1}$ can be satisfied for $\pi/4\ll|\omega_1 t|\leq \pi/2$, where $1/2\ll {\rm Pr}_{\pm\to \mp}^{\rm Rabi}$ holds and the transition probability is significantly enhanced.
This shows that the Rabi resonance involving significant enhancement $1/2\ll{\rm Pr}_{\pm\to \mp}^{\rm Rabi}$ can be considered as the weak value amplification with the measurement strength $\delta_{\rm Rabi}$.
In particular, for $\omega_1 t=\pi/2$, the Rabi resonance condition corresponds to the divergence condition of the weak value, see Eq.~\eqref{eq:weaksRab}.
%

From Eq.~\eqref{eq:Prr}, the imaginary component of the weak value of $\sigma_2$ is obtained as
\begin{align}
    {\rm Im}\,\sigma^W_{2,L}(\phi_{\rm Rabi},\omega_1 t)=\frac{1}{{\rm Pr}^{\rm Rabi}_{\pm\to\pm}(0)}\frac{d{\rm Pr}^{\rm Rabi}_{\pm\to\pm}(\epsilon)}{d\delta_{\rm Rabi}}\bigg|_{\delta_{\rm Rabi}=0},\label{eq:weakrab}
\end{align}
which can be evaluated from the measured values of $\delta_{\rm Rabi}$.
Eq.~\eqref{eq:weakrab} implies that the measurement of $\epsilon$ based on the Rabi resonance potentially determines the imaginary component of the weak value of $\sigma_2$.
As mentioned below Eq.~\eqref{eq:pr1}, the non-unitary time evolution is required to  extract the real part of the weak values.
This is why the imaginary part of only the weak value can be obtained in the resonance phenomenon.
Since the weak value of a physical quantity is defined in the limit of zero measurement strength, it is generally understood as the physical quantity possessed by the system when no disturbance is inflicted. 
The detail of experiments determining the weak value will be discussed in Sec.~\ref{sec:4} in the context of the Ramsey resonance.

We recall that the weak value amplification can be achieved by adjusting the pre- and post-selected states, ${|\psi_i\rangle}$ and ${|\psi_f\rangle}$, in such a way that the denominators of Eq.~\eqref{eq:weakR} reduce to zero.
Since in the present case ${|\psi_i(\phi_{\rm Rabi})\rangle}$ and ${|\psi_f(\phi_{\rm Rabi})\rangle}$ contain only the parameter $\phi_{\rm Rabi}=(\omega-\overline{\omega}_0)/2\omega_1$ defined in a single time region of $t_0<t$, the resonance condition $\omega=\omega_0$ is specified by the parameters in the single time region.  It is, however, possible to alter the resonance condition by including other parameters associated with the pre-selected and post-selected states, enlarging the scope of parameters which affect the resonance condition.  Indeed, the Ramsey resonance condition discussed below utilizes such parameters other than those in the original time region.
%

\subsection{Ramsey resonance}
\label{sec:Ramsey}
%
In order to discuss the Ramsey resonance, in the context of weak measurement, let us consider the Hamiltonian describing the dynamics of a particle spin separated in three time regions,
\begin{widetext}
\begin{align}
    H_{\rm Ramsey}(t):=\begin{cases}
    -\frac{\omega_0}{2}\sigma_3+\omega_2 \left(\cos\omega t \sigma_1-\sin\omega t\sigma_2\right),~~&t_0\leq t < t_0+\tau/2,
    \\
    -\frac{\omega_0}{2}\sigma_3,~~&t_0+\tau/2\leq t < t_0+\tau/2+T,
    \\
    -\frac{\omega_0}{2}\sigma_3+\omega_2\left(\cos\omega t \sigma_1-\sin\omega t\sigma_2\right),~~&t_0+\tau/2+T\leq t < t_0+\tau+T.
    \end{cases}
    \label{eq:HRamsey1}
\end{align}
\end{widetext}
For brevity, we consider a typical case of Ramsey resonance, focusing on $\omega_0=\omega$ in Eq.~\eqref{eq:HRamsey1}. 
Even without this condition $\omega_0=\omega$, the following explanations are applicable for $\tau\ll T$.
Since the Hamiltonian of the first and third time regions is the same as $H_{\rm Rabi}$ in Eq.~\eqref{eq:hamRabi} with $\omega_1=\omega_2$, from Eq.~\eqref{eq:timeIII} the time evolution of the initial state $|\psi(t_0)\rangle$ is given by
%
%
\begin{widetext}
\begin{align}
&{|\psi(t_0+\tau/2)\rangle}=e^{i\omega (t_0+\tau/2)\sigma_3/2}e^{-i \omega_2\tau\sigma_1/2}e^{-i \omega t_0\sigma_3/2}{|\psi(t_0)\rangle},\label{eq:UN1}
\\
&{|\psi(t_0+\tau/2+T)\rangle}=e^{i\omega_0 \sigma_3 T/2}{|\psi(t_0+\tau/2)\rangle},\label{eq:UN2}
\\
&{|\psi(t_0+\tau+T)\rangle}=e^{i\omega (t_0+\tau+T) \sigma_3/2} e^{-i\omega_2 \tau\sigma_1/2}e^{-i \omega(t_0+\tau/2+T)\sigma_3/2}{|\psi(t_0+\tau/2+T)\rangle}.\label{eq:UN3}
\end{align}
Combining Eqs.~\eqref{eq:UN1}-\eqref{eq:UN3}, we find the final state,
\begin{align}
    {|\psi(t_0+\tau+T)\rangle}=e^{i\omega (t_0+\tau+T)\sigma_3/2}e^{-i\omega_2 \tau \sigma_1/2}e^{-i (\omega-\omega_0) T\sigma_3/2} e^{-i \omega_2\tau\sigma_1/2}e^{-i \omega t_0\sigma_3/2}{|\psi(t_0)\rangle}.\label{eq:timramsey}
\end{align}
For our later convenience, we combine all unitary operators appearing in the time evolution into one:
\begin{align}
U(t_0,t_0+\tau+T):=e^{i\omega (t_0+\tau+T)\sigma_3/2}e^{-i\omega_2 \tau \sigma_1/2}e^{-i (\omega-\omega_0) T\sigma_3/2} e^{-i \omega_2\tau\sigma_1/2}e^{-i \omega t_0\sigma_3/2}.\label{eq:uniRamsey}
\end{align}

Now, as in the case of the Rabi resonance, let us consider the initial state ${|\psi(t_0)\rangle}={|\pm\rangle}$.
From Eq.~\eqref{eq:timramsey}, the transition amplitude from ${|\pm\rangle}$ to ${|\pm\rangle}$ is obtained as
\begin{align}
    {\langle \pm|\psi(t_0+\tau+T)\rangle}&=e^{\pm i\omega (\tau+T)/2} {\langle \pm|}e^{-i\omega_2 \tau \sigma_1/2}e^{-i(\omega-\omega_0)T \sigma_3/2}e^{-i\omega_2\tau \sigma_1/2}{|\pm\rangle}.\label{eq:ampRamsey}
\end{align}
\end{widetext}
From Eq.~\eqref{eq:ampRamsey}, the transition probability then reads
\begin{align}
    {\rm Pr}_{\pm\to\pm}^{\rm Ramsey}=\left|{\langle \pm|}e^{-i\omega_2 \tau \sigma_1/2}e^{-i(\omega-\omega_0)T \sigma_3/2}e^{-i\omega_2\tau \sigma_1/2}{|\pm\rangle}\right|^2.\label{eq:prRamsey}
\end{align}
As before, we further consider the case $\omega=\omega_0$ for which we obtain
\begin{align}
    {\rm Pr}_{\pm\to\pm}^{\rm Ramsey}=\left|{\langle \pm|}e^{-i\omega_2 \tau \sigma_1}{|\pm\rangle}\right|^2~{\rm for}~\omega=\omega_0.~\label{eq:proIIIramsey}
\end{align}
Eq.~\eqref{eq:proIIIramsey} shares the same form with Eq.~\eqref{eq:proIII2}, and in particular, for $\omega_2\tau=\pi/2$ the probability becomes zero.  This indicates a resonance to arise there, which is referred to as the Ramsey resonance.
It should be emphasized, however, that the Ramsey resonance is inherently different from the Rabi resonance.  In fact, 
while the Rabi resonance arises between two oscillations $\omega$ and $\omega_0$ in the same time region, 
the Ramsey resonance arises between two oscillations $\omega$ and $\omega_0$ observed in different time regions.

Next, we discuss the Ramsey resonance from the viewpoint of weak measurement.
To this end, similarly to Eq.~\eqref{eq:ep1} we write the parameter $\omega_0$ as $\omega_0:=\overline{\omega}_0+\epsilon$ and consider the weak measurement of $\epsilon$ in the following setup:  
\begin{itembox}[l]{Weak measurement via Ramsey resonance}
\begin{itemize}
    \item Known parameters: $\overline{\omega}_0$, $\omega$, $\omega_2$, $\tau$, and $T$
    \item Unknown parameter: $\epsilon$
    \item Condition of weak measurement: $\epsilon T\ll1$
\end{itemize}
\end{itembox}
%
%
Then, for $t_0+\tau/2\leq t <t_0+\tau/2+T$, the corresponding Hamiltonian in \eqref{eq:HRamsey1} is rewritten as
\begin{align}
    H_{\rm Ramsey}(t)=-\frac{\overline{\omega}_0}{2}\sigma_3-\frac{\epsilon}{2}\sigma_3.\label{eq:ramsec}
\end{align}
According to the above setup, we choose the two operators $H_0$ and $V$ in  \eqref{eq:Ham1} as
\begin{align}
    H_0 =-\frac{\overline{\omega}_0}{2}\sigma_3,~~~V=-\frac{\epsilon}{2} \sigma_3.
\end{align}
and also the coefficients appearing in Eq.~\eqref{eq:h0V}  as
\begin{align}
    &h:=\frac{\overline{\omega}_0}{2},~~n^{(h)}_0:=0,~~\vec{n}^{(h)}=\left(0,0,-1\right),\notag
    \\   &v:=\frac{\epsilon}{2},~~n_0^{(v)}:=0,~~\vec{n}^{(v)}:=\left(0,0,-1\right).
\end{align}
Since $\vec{n}^{(h)}\times \vec{n}^{(v)}=0$ holds, we are in the commutative case mentioned in Sec.~\ref{sec:2}.
If we choose ${|\psi_i\rangle}:=e^{i\omega (t_0+\tau/2)\sigma_3/2}e^{-i\omega_2\tau\sigma_1/2}e^{-i\omega t_0\sigma_3/2}{|\pm\rangle}$, and ${|\psi_f\rangle}=e^{i\omega (t_0+\tau/2+T)\sigma_3/2}e^{i\omega_2\tau\sigma_1/2}e^{-i\omega(t_0+\tau+T)\sigma_3/2}{|\pm\rangle}$, then from Eqs.~\eqref{eq:PrI} and \eqref{eq:ramsec} the transition probability from ${|\psi_i\rangle}$ to ${|\psi_f\rangle}$ is given by
\begin{align}
    {\rm Pr}_{\pm\to \pm}^{\rm Ramsey}(\epsilon)\simeq {\rm Pr}_{\pm\to \pm}^{\rm Ramsey}(0) \left(1- \epsilon T\,{\rm Im}\, \sigma_3^W\right),
\end{align}
where
\begin{align}
    {\rm Pr}_{\pm\to \pm}^{\rm Ramsey}(\epsilon):=\left|{\langle \psi_f| e^{-i\int_{t_0+\tau/2}^{t_0+\tau/2+T}dt H_{\rm Ramsey}(t)}|\psi_i\rangle} \right|^2,\label{eq:PrRamseyif}
\end{align}
and
\begin{align}
    \sigma_3^W =\frac{{\langle \psi_f|e^{-i H_0 T}\sigma_3|\psi_i\rangle}}{{\langle \psi_f| e^{-i H_0 T}|\psi_i\rangle}}=\frac{{\langle \psi_f|\sigma_3e^{-i H_0 T}|\psi_i\rangle}}{{\langle \psi_f| e^{-i H_0 T}|\psi_i\rangle}}.\label{eq:weak3}
\end{align}

In particular, for $\omega_2\tau=\pi/2$, the following relations hold:
\begin{widetext}
\begin{align}
    &{\langle \psi_f|e^{-i H_0 T }\sigma_3|\psi_i\rangle}=e^{\pm i\omega(\tau+T)/2}{\langle \psi_f(\phi_{\rm Ramsey})| e^{-i\omega_2\tau \sigma_1}\sigma_2 |\psi_i(\phi_{\rm Ramsey})\rangle},\label{eq:we1p}
\\
&{\langle \psi_f|e^{-i H_0 T }|\psi_i\rangle}=e^{\pm i\omega(\tau+T)/2}{\langle \psi_f(\phi_{\rm Ramsey})|e^{-i\omega_2\tau \sigma_1} |\psi_i(\phi_{\rm Ramsey})\rangle},\label{eq:we2p}
\end{align}
\end{widetext}
where
\begin{align}
{|\psi_i(\phi_{\rm Ramsey})\rangle}={|\psi_f(\phi_{\rm Ramsey})\rangle}:=e^{-i\phi_{\rm Ramsey}\, \sigma_2/2}{|\pm\rangle},
\end{align}
with 
\begin{align}
\phi_{\rm Ramsey}:=(\omega-\overline{\omega}_0)\frac{T}{2}
   \label{eq:Ramseyphase}
\end{align}
From Eqs.~\eqref{eq:we1p} and \eqref{eq:we2p}, for $\omega_2 \tau=\pi/2$, we find the weak value,
\begin{align}
    \sigma_3^W &=- \sigma_{2,L}^W\left(\phi_{\rm Ramsey},\pi/2\right)= \sigma_{2,R}^W\left(\phi_{\rm Ramsey},\pi/2\right)\notag
    \\
    &= i\cot \phi_{\rm Ramsey}.\label{eq:weak3Ramsey}
\end{align}
The imaginary parts of the weak values are generally given as,
\begin{align}
    &{\rm Im}\,\sigma_3^W=- {\rm Im}\,\sigma_{2,L}^W\left(\phi_{\rm Ramsey},\omega_2\tau\right)= {\rm Im}\,\sigma_{2,R}^W\left(\phi_{\rm Ramsey},\omega_2\tau\right)\notag
    \\
    &\quad\quad\quad= \cot \phi_{\rm Ramsey}\times \frac{\sin^2 \phi_{\rm Ramsey} \tan^2 (\omega_2\tau)}{1+\sin^2 \phi_{\rm Ramsey} \tan^2 (\omega_2\tau)}.
\end{align}
Also, the transition probability for $\epsilon=0$ is generally evaluated as,
\begin{align}
    {\rm Pr}_{\pm\to \pm}^{\rm Ramsey}(0)&=\left|{\langle \psi_f|e^{+i \overline{\omega}_0 T \sigma_3/2} |\psi_i\rangle}\right|^2\notag
    \\
    &=\sin^2\phi_{\rm Ramsey}\left(1+\cos^2(\omega_2\tau) \cot^2\phi_{\rm Ramsey} \right).
\end{align}
This expression hold even for $\epsilon\neq 0$ by replacing $\phi_{\rm Ramsey}\to \phi_{\rm Ramsey}-\epsilon T/2$ because the shift from the resonance point $\epsilon$ can be absorbed to $\phi_{\rm Ramsey}$.
With these, Eq.~\eqref{eq:PrRamseyif} is evaluated as
\begin{align}
    {\rm Pr}_{\pm\to \pm}^{\rm Ramsey}(\epsilon)\simeq {\rm Pr}_{\pm\to \pm}^{\rm Ramsey}(0) \left(1+\delta_{\rm Ramsey}\,{\rm Im}\,\sigma^W_{2,L}\left(\phi_{\rm Ramsey},\omega_2\tau\right) \right), \label{eq:proramseyf}
\end{align}
where $\delta_{\rm Ramsey}:=\epsilon T$ is the measurement strength for the Ramsey resonance, and ${\rm Im}\,\sigma^W_{2,L}$ is given in Eq.~\eqref{eq:Rabi_weak_im}.
Similarly to the Rabi resonance, in the limit $\phi_{\rm Ramsey}=0$ under the condition $\tan^2 (\omega_2\tau)\geq |\phi_{\rm Ramsey}|^{-1}$, the probability \eqref{eq:proramseyf} is significantly suppressed and the weak value is simultaneously amplified so that $|{\rm Im}\,\sigma^W_{2,L}\left(\phi_{\rm Ramsey},\omega_2\tau\right)|=|{\rm Im}\,\sigma^W_{2,R}\left(\phi_{\rm Ramsey},\omega_2\tau\right)|\geq 1$.
In particular, for $\omega_2 \tau=\pi/2$, the probability \eqref{eq:proramseyf} becomes zero and the weak value \eqref{eq:weak3Ramsey} diverges.
This indicates that the Ramsey resonance involving significant enhancement $1/2\ll{\rm Pr}^{\rm Ramsey}_{\pm\to \mp}$ can be regarded as weak value amplification with the measurement strength $\delta_{\rm Ramsey}$. 

It is notable that, except for the magnitude of measurement strength, the two probability formulae, Eqs.~\eqref{eq:Prr} and \eqref{eq:proramseyf}, become identical at $T=t$ and $\omega_1 t=\omega_2 \tau$.
In particular, for $\omega_1 t=\omega_2 \tau$, we obtain $\phi_{\rm Ramsey}=\phi_{\rm Rabi}$ as well as $\delta_{\rm Ramsey}=\delta_{\rm Rabi}$, if we replace $T$ with $1/\omega_1$.
This implies that the weak measurement offers a unified framework for resonances in inherently different types, one being commutative and the other non-commutative in the classification given in Sec.~\ref{sec:2}.  It also shows that the measurement strength presents a means to quantify how sensitive the resonance is against small changes in resonance conditions.
Such parallel becomes available due to our different choices of the initial and final states, ${|\psi_i\rangle}$ and ${|\psi_f\rangle}$,  for the Ramsey and Rabi resonances. 

We also note that from Eq.~\eqref{eq:proramseyf} the imaginary component of the weak value is obtained as
\begin{align}
    {\rm Im}\,\sigma^W_{2,L}(\phi_{\rm Ramsey},\omega_2\tau)=\frac{1}{{\rm Pr}_{\pm\to\pm}^{\rm Ramsey}(0)}
    \frac{d {\rm Pr}_{\pm\to\pm}^{\rm Ramsey}(\epsilon)}{d\delta_{\rm Ramsey}}\bigg|_{\delta_{\rm Ramsey}=0}.\label{eq:weakRams0}
\end{align}
As in the case of the Rabi resonance, the right-hand side of Eq.~\eqref{eq:weakRams0} can be evaluated from the measurement of $\delta_{\rm Ramsey}$, that is, the Ramsey resonance determines the imaginary component of the weak value of $\sigma_2$.
Based on these results, we discuss the experiments determining the weak value in more detail in Sec.~\ref{sec:4} by taking  some realistic factors into consideration.

\subsection{Comparison between Rabi and Ramsey resonances in light of weak measurement}
Before doing so, we revisit the Rabi and Ramsey resonances and compare them in light of the weak measurement.
Recall that we have considered the case where the resonance condition is slightly broken by the unknown parameter $\epsilon$ and regarded that the disturbance of the system by $\epsilon$ is required in the weak measurement process. 
For our purpose, we may consider the particular choice $t=T$, which is allowed because the Ramsey resonance often assumes $\tau\ll T$ to ensure insensitivity to inhomogeneities of external electric/magnetic fields. 
Our comparison is based on the three key properties of resonance we found so far:
%
%
%

    \noindent
    (i) {\it Resonance as weak value amplification} ---
    Both the Rabi and Ramsey resonances are regarded as weak value amplification, and share a common feature that their weak values \eqref{eq:weaksRab} and \eqref{eq:weak3Ramsey} diverge as
    \begin{align}
    \lim_{\phi\to 0}{\rm Im}\,\sigma^W_{2,L}\left(\phi,\pi/2\right)=-\lim_{\phi\to 0}{\rm Im}\,\sigma^W_{2,R}\left(\phi,\pi/2\right)=-\infty,
    \end{align}
    for $\phi = \phi_{\rm Rabi}$ or $\phi_{\rm Ramsey}$, when $\omega_1 t= \pi/2$ (Rabi resonance) and $\omega_2 \tau=\pi/2$ (Ramsey resonance) hold.
    These resonances are regarded as weak value amplification even for $\omega_1 t\neq \pi/2$ (Rabi resonance) and $\omega_2 \tau\neq\pi/2$ (Ramsey resonance) when the significant enhancement of the probabilities $1/2\ll{\rm Pr}^{\rm Rabi/Ramsey}_{\pm\to\mp}$ is achieved.
    It is worth mentioning that $\phi_{\rm Rabi}$ and $\phi_{\rm Ramsey}$ in the resonance phenomena represent parameters characterizing the pre- and post-selected states.
    
    \noindent
    (ii) {\it Probability near resonance} ---
    For $t=T$ and $\omega\simeq \overline{\omega}_0$, the formula for the transition probability from ${|\pm\rangle}$ to ${|\pm\rangle}$ is formally identical for the Rabi and Ramsey resonances:
    \begin{widetext}
    \begin{align}
    &{\rm Pr}_{\pm\to \pm}^{\rm Rabi}(\epsilon)\simeq {\rm Pr}_{\pm\to \pm}^{\rm Rabi}(0)\left(1+\delta_{\rm Rabi}\,{\rm Im}\,\sigma^W_{2,L}\left(\phi_{\rm Rabi},\omega_1 t\right)\right)\quad[{\rm Rabi~resonance}],\label{eq:rabiLA}
    \\
    &{\rm Pr}_{\pm\to \pm}^{\rm Ramsey}(\epsilon)\simeq {\rm Pr}_{\pm\to \pm}^{\rm Ramsey}(0)\left(1+\delta_{\rm Ramsey}\, {\rm Im}\,\sigma^W_{2,L}\left(\phi_{\rm Ramsey},\omega_2 \tau\right)\right)\quad[{\rm Ramsey~resonance}],\label{eq:ramiLA}
    \end{align}
    \end{widetext}
with $\delta_{\rm Rabi}=\epsilon /\omega_1$, $\delta_{\rm Ramsey}=\epsilon T$ and $\phi_{\rm Rabi}=(\omega-\overline{\omega}_0)/2\omega_1$, $\phi_{\rm Ramsey}=(\omega-\overline{\omega}_0)T/2$ as given in \eqref{eq:Rabiphase} and \eqref{eq:Ramseyphase}.  We have then ${\rm Pr}_{\pm\to\pm}^{\rm Rabi}(0)=\sin^2\phi_{\rm Rabi}\left(1+\cos^2(\omega_1 t)\cot^2\phi_{\rm Rabi}\right)$ and ${\rm Pr}_{\pm\to\pm}^{\rm Ramsey}(0)=\sin^2\phi_{\rm Ramsey}\left(1+\cos^2 (\omega_2\tau)\cos^2\phi_{\rm Ramsey}\right)$.
    For $t=T$, $\omega_1 t=\omega_2 \tau$, and $\phi_{\rm Rabi}=\phi_{\rm Ramsey}=\phi$, the difference between Eqs.~\eqref{eq:rabiLA} and \eqref{eq:ramiLA} arises only from the difference between $\delta_{\rm Rabi}$ and $\delta_{\rm Ramsey}$. 

    \noindent
    (iii) {\it Measurement strength in resonance} ---
    For $t=T$, the measurement strengths of the Rabi and Ramsey resonances are $\delta_{\rm Rabi}=\epsilon /\omega_1$ and $\delta_{\rm Ramsey}=\epsilon t$, respectively.
    In particular, $\pi/4t \ll|\omega_1| \leq\pi/2t$ holds for the Rabi resonance involving the significant enhancement of the transition probability, {\it i.e.}, $1/2\ll{\rm Pr}^{\rm Rabi}_{\pm\to\mp}$.
    Given this, at most, the measurement strength of the Ramsey resonance can be $\pi/2\simeq 1/0.6$ times stronger than that of the Rabi resonance, implying that the Ramsey resonance can have higher sensitivity to the shift of the resonance condition than the Rabi resonance.
%

From (i), (ii), and (iii), we see that the perspective of the weak measurement provides a unified understanding of the Rabi and Ramsey resonances where the sensitivity to the shift of the resonance condition is characterized by the measurement strength, in spite of the fact that the two resonances fall in different types of weak measurements, one being commutative and the other non-commutative.

\section{Weak value in neutron EDM experiment}
\label{sec:4}
%
We have seen that the resonance involving significant enhancement of the transition probability can generally be regarded as weak value amplification.  Given this, we now 
propose a weak value measurement of neutron EDM.
In what follows, we assume $\omega_2\tau=\pi/2$ in the Ramsey resonance because the neutron EDM experiment concentrates on this case.
\subsection{Imperfection effects of pre- and post-selections on Ramsey resonance}
In preparation for the discussion of neutron EDM measurement, we will expand on the Ramsey resonance in Sect.~\ref{sec:Ramsey} by taking realistic factors into consideration, that is, the possible imperfection of pre- and post-selection that arise in actual systems possessing internal structures under the influence of environment.
For this, rather than the pure states considered in Sect.~\ref{sec:Ramsey}, we consider a mixed state for the initial state of the system,
\begin{align}
    \rho_{i}:=\frac{1+P_i}{2}{|+\rangle}{\langle +|}+ \frac{1-P_i}{2} {|-\rangle}{\langle -|},\label{eq:initialNeut}
\end{align}
where $P_i$ denotes the statistical disparity of spin characterizing the imperfection of the pre-selected state.
We also consider the imperfection of post-selection, which is dealt with the POVM operator:
\begin{align}
    E_+:= (1-\epsilon_f){|+\rangle}{\langle +|}+\epsilon_f{|-\rangle}{\langle -|},\label{eq:me1}
    \\
    E_-:=\epsilon_f {|+\rangle}{\langle +|}+(1-\epsilon_f){|-\rangle}{\langle -|},\label{eq:me2}
\end{align}
where $E_+$ and $E_-$ correspond to the spin-up and spin-down states, respectively, and $\epsilon_f$ represents the imperfection of the spin observations. 

Recall that the time evolution consists of the three time-regions, and the unitary time evolution is dictated by Eq.~\eqref{eq:uniRamsey} for $\omega_0=\overline{\omega}_0+\epsilon$.   The probability that the spin-up or -down state is observed at $t_0+\tau+T$ is then found to be
\begin{widetext}
\begin{align}
    {\rm Pr}_{i\to f}^{\rm Ramsey}(\pm;\epsilon)&={\rm Tr}\left[
    E_{\pm}U(t_0,t_0+\tau+T)\rho_iU^{\dagger}(t_0,t_0+\tau+T)
    \right]\notag
    \\
    &=\frac{1+P_i}{2}\left(
    (1-\epsilon_f) {\rm Pr}_{+\to \pm}^{\rm Ramsey}(\epsilon)+\epsilon_f {\rm Pr}_{+\to \mp}^{\rm Ramsey} (\epsilon)
    \right)+\frac{1-P_i}{2}\left(
    (1-\epsilon_f) {\rm Pr}_{-\to \pm}^{\rm Ramsey}(\epsilon)+
    \epsilon_f {\rm Pr}_{-\to \mp}^{\rm Ramsey}(\epsilon)
    \right),\label{eq:proReal}
\end{align}
where
\begin{align}
    {\rm Pr}_{+\to -}^{\rm Ramsey}(\epsilon)&=\left|{\langle - |U(t_0,t_0+\tau+T)|+\rangle}\right|^2=\frac{1}{2}\left(1+\cos \left(2\phi_{\rm Ramsey}-\epsilon T\right)\right)~{\rm for}~\omega_2\tau=\frac{\pi}{2},\label{eq:com1}
    \\
    {\rm Pr}_{+\to +}^{\rm Ramsey}(\epsilon)&=\left|{\langle + |U(t_0,t_0+\tau+T)|+\rangle}\right|^2=\frac{1}{2}\left(1-\cos \left(2\phi_{\rm Ramsey}-\epsilon T\right)\right)~{\rm for}~\omega_2\tau=\frac{\pi}{2}
    ,
    \\
    {\rm Pr}_{-\to -}^{\rm Ramsey}(\epsilon)&=\left|{\langle - |U(t_0,t_0+\tau+T)|-\rangle}\right|^2=\frac{1}{2}\left(1-\cos \left(2\phi_{\rm Ramsey}-\epsilon T\right)\right)~{\rm for}~\omega_2\tau=\frac{\pi}{2}
    ,
    \\ 
    {\rm Pr}_{-\to +}^{\rm Ramsey}(\epsilon)&=\left|{\langle + |U(t_0,t_0+\tau+T)|-\rangle}\right|^2=\frac{1}{2}\left(1+\cos \left(2\phi_{\rm Ramsey}-\epsilon T\right)\right)~{\rm for}~\omega_2\tau=\frac{\pi}{2},
\end{align}
with $\phi_{\rm Ramsey}$ given in \eqref{eq:Ramseyphase}.
Up to the first order of $\epsilon$, the above probabilities for $\omega_2\tau=\pi/2$ can be expressed by the weak values as
\begin{align}
    {\rm Pr}_{+\to-}^{\rm Ramsey}(\epsilon)&={\rm Pr}_{-\to+}^{\rm Ramsey}(\epsilon)\simeq \left(1-{\rm Pr}_{+\to+}^{\rm Ramsey}(0)\right)\left(1+\epsilon T \frac{{\rm Pr}_{+\to +}^{\rm Ramsey}(0)}{1-{\rm Pr}_{+\to +}^{\rm Ramsey}(0)}\,{\rm Im}\,\sigma^{W}_{3} \right), 
    \\
    {\rm Pr}_{+\to +}^{\rm Ramsey}(\epsilon)&={\rm Pr}_{-\to -}^{\rm Ramsey}(\epsilon)\simeq {\rm Pr}_{+\to +}^{\rm Ramsey}(0)\left(1-\epsilon T\,{\rm Im}\,\sigma^{W}_{3}\right),\label{eq:com2}
\end{align}
with the weak value $\sigma^W_3$ given in Eq.~\eqref{eq:weak3}.
Substituting Eqs.~\eqref{eq:com1}-\eqref{eq:com2} into Eq.~\eqref{eq:proReal}, for $\omega_2 \tau=\pi/2$, we obtain
\begin{align}
    {\rm Pr}^{\rm Ramsey}_{i\to f}\left(\pm;\epsilon\right)&=\frac{1}{2}\left(1\mp \alpha \cos\left(2\phi_{\rm Ramsey}-\epsilon T\right)\right)\label{eq:proramweak0}
    \\
    &\simeq \frac{1}{2}\left[
    1\mp \alpha \left(
    1-2{\rm Pr}^{\rm Ramsey}_{+\to +}\left(0\right)+2\epsilon T\,{\rm Pr}^{\rm Ramsey}_{+\to +}\left(0\right)\,{\rm Im}\,\sigma^W_3
    \right)
\right],\label{eq:proramweak}
\end{align}
where 
\begin{align}
\alpha:=P_i (1-2\epsilon_f)
\label{eq:alphapara}
\end{align}
represents the overall degree of perfection in state preparation and measurement currently considered.  
From Eqs.~\eqref{eq:proramweak} and \eqref{eq:proramweak}, we learn that in actual systems under the influence of environmental noise, the sensitivity of the resonance to $\epsilon$ is proportional to $\alpha$ in addition to time $T$.  Since $\alpha$ in \eqref{eq:alphapara} represents the combined effects of imperfections in pre- and post-selections and vanishes $\alpha = 0$ for the most erroneous state selections, $P_i = 0$ or $\epsilon_f = 1/2$, we observe that the merit of resonance is tempered according to the degrees of imperfections even at the resonance points as expected.
\end{widetext}

From Eq.~\eqref{eq:proramweak}, the imaginary part of the weak value can be retrieved as
\begin{align}
{\rm Im}\,\sigma^W_3=\frac{2}{1\mp \alpha-2 {\rm Pr}_{i\to f}^{\rm Ramsey}\left(\pm;0\right)}\frac{d {\rm Pr}_{i\to f}^{\rm Ramsey}\left(\pm;\epsilon\right)}{d(\epsilon T)}\bigg|_{\epsilon=0}.\label{eq:weakRams}
\end{align}
Since the right-hand side of Eq.~\eqref{eq:weakRams} consists of measurable quantities, we realize that the weak value ${\rm Im}\,\sigma^W_{3}$ can be determined through experiment realizing the Ramsey resonance.
Below, we discuss a neutron EDM measurement in which ${\rm Im}\,\sigma^W_3$ can be obtained using the Ramsey resonance technique even when the presence of imperfection cannot be ignored.

%
%
%

\subsection{Neutron EDM measurement experiment}

In the neutron EDM measurement, the above Ramsey resonance technique has actually been widely applied, where the parameters in Eqs.~\eqref{eq:ep1} and \eqref{eq:HRamsey1} correspond to
\begin{align}
    \overline{\omega}_0= 2\mu_{\rm n} B_0,~\epsilon=2 d_{\rm n} E_0,~\omega_2=\mu_{\rm n} B _1/2.
\end{align}
Here, $\mu_{\rm n}$ is the magnetic dipole moment of the neutron, $d_{\rm n}$ is the EDM of the neutron, $B_0$ and $E_0$ are static magnetic and electric fiedls, respectively, and $B_1$ denotes the magnitude of rotating magnetic field.
Currently, the value of the neutron EDM has not been determined through experiments, and only its upper limit has been measured~\cite{Abel:2020pzs}.
Namely, to date, $\epsilon$ is not inconsistent with zero in the experiments.
We follow Refs.~\cite{Baker:2006ts,Pendlebury:2015lrz} and briefly review some of the key points on technical descriptions of the experiment.

Most recent experiments on neutron EDM use ultra-cold neutrons (UCNs) which move at a velocity of less than 7 m/s. 
At the start of the Ramsey resonance technique, the neutrons pass through the magnetized polarizing foil in order to prepare
the initial state \eqref{eq:initialNeut}, and the tranmitted neutrons are stored in an evacuated chamber that has walls reflecting the neutrons.
The rotating magnetic field $B_1$, perpendicular to $B_0$, is applied during the interval $\tau/2= 2\,$s.
Subsequently, we let the neutrons precess during the interval $T= 130\,$s until the second rotating magnetic field $B_1$.
After this, the neutrons are released from the evacuated chamber, and the number of neutrons with spin up (and those with spin down) is counted by using the magnetized polarizing foil, which implements the measurement of Eqs.~\eqref{eq:me1} and \eqref{eq:me2}. 

In one of the measurements this cycle was repeated many times, and each cycle yielded about 14000 UCNs.  Namely, these are obtained after cuts were applied to data gained in 545 runs containing 175,217 measurement cycles with $2.5 \times 10^9$ neutrons remained~\cite{Pendlebury:2015lrz}. 
It is apparent from Eq.~\eqref{eq:proramweak0} that to measure the EDM effect $\epsilon$, the magnetic field $B_0$ must be accurately controlled.
A $^{199}$Hg cohabiting magnetometer is utilized to attain a precise magnetic field, that is, polarized $^{199}$Hg vapor is filled with the same chamber as the UCNs and the magnetic field $B_0$ is measured by monitoring the mercury precession frequency $\omega_{\rm Hg}$.
By using the cohabiting magnetometer, the first-order estimate of the neutron precession frequency is determined as $\omega_{\rm n,Hg}:=|\mu_{\rm n}/\mu_{\rm Hg}|\omega_{\rm Hg}$ where $\mu_{\rm Hg}$ is the magnetic dipole moment of the mercury.

Other experiments have precisely measured the ratio of magnetic dipole moments, which has a known value $\mu_{\rm n}/\mu_{\rm Hg}=-3.8424574(30)$~\cite{Afach:2014fha}.
From Eq.~\eqref{eq:proramweak0}, the number of counted spin-up or down neutrons $N_\pm$ reads~\footnote{The dark counts may occur due, {\it e.g.}, to possible neutron decay. 
Such effects will be included in the prefactor $\overline{N}$.}
\begin{align}
    N_{\pm}=\overline{N}\left(
    1\mp \alpha \cos\left(2\phi_{\rm Ramsey}-\epsilon T \right)
    \right),\label{eq:Npm}
\end{align}
in which we have
\begin{align}
    2\phi_{\rm Ramsey}-\epsilon T&=\left(\omega-\omega_0\right) T\notag
    \\
    &=\left[\left(\omega-\omega_{\rm n,Hg}\right)-\left(\omega_0-\omega_{\rm n,Hg}\right)\right] T\notag
    \\
    &=\left(\delta \omega -\Phi\right) T,
\end{align}
where $\delta \omega:=\omega-\omega_{\rm n,Hg}$ and $\Phi:=\omega_0-\omega_{\rm n,Hg}$.
%
The true neutron frequency $\omega_0$ may differ from the first-order estimate $\omega_{\rm n,Hg}$ measured from the mercury cohabiting magnetometer for several reasons, such as the EDM effect, the inherent uncertainty in the ratio $\mu_{\rm n}/\mu_{\rm Hg}$, and the difference in the spatial distribution in the chamber between UCNs and $^{199}$Hg via gravity.
The above measurements were conducted by tuning $\delta\omega$ and varying the directions of the electric and magnetic fields depending on the cycles, and each run, typically consisting of several hundred measurement cycles, contains the cycles with different $\delta\omega$ and directions of the fields.
If the electric and magnetic fields are parallel or anti-parallel, the precession frequency $\omega_0$ is expressed as $\omega_0=-2\mu_{\rm n} |B_0|\mp 2d_{\rm n}|E_0|$.
In the approach of Ref.~\cite{Pendlebury:2015lrz}, for each run, the neutron counts $N_{\pm}$ were fitted to Eq.~\eqref{eq:Npm} for each of spin-up and -down states, and thereby determined $\overline{N}, \alpha$ and $\Phi$ for each of the two spin states.
Considering this, for each run, Eq.~\eqref{eq:Npm} is now be written as
\begin{align}
    N_{\pm}=\overline{N}_{\pm} \left(1\mp \alpha_{\pm} \cos\left(\delta \omega -\Phi_{\pm}\right)T\right).\label{eq:runN}
\end{align}
Note here that $\Phi_{\pm}$ is averaged over each run, during which the effects of EDM are expected to cancel out.
In contrast, for each measurement cycle $j$ and for each spin state, Eq.~\eqref{eq:Npm} should become
\begin{align}
    N_{\pm,j}=\overline{N}_{\pm} \left(1\mp \alpha_{\pm} \cos\left(\delta \omega -\Phi_{\pm,j}\right)T\right).\label{eq:cycleN}
\end{align}
Comparing Eqs.~\eqref{eq:runN} and \eqref{eq:cycleN}, we find that the EDM contribution to the measurement cycle $j$ may be given by 
\begin{align}
    \epsilon_{\pm,j}:&=\Phi_{\pm,j}-\Phi_{\pm}\notag
    \\
    &=\delta \omega -\Phi_{\pm}- \frac{1}{T}\arccos\left[\frac{N_{\pm,j}-\overline{N}_{\pm}}{\overline{N}_{\pm} \alpha_{\pm}}\right].
\end{align}

Meanwhile, we notice from Eq.~\eqref{eq:Npm} that 
\begin{align}
    \frac{d N_{\pm}}{d \omega}=\pm \overline{N} \alpha T \sin \left(\omega-\omega_0\right)T, 
    \label{eq:delN}
\end{align}
which implies that measurements at $ \omega\simeq \omega_0\pm \pi/2T$ possess the maximal sensitivity for $N_{\pm}$ against $\epsilon$. 
In the neutron EDM experiment, the statistical uncertainty has been known to dominate over the systematic uncertainty.
Since the fractional uncertainty in the number of neutrons counted is at best $1/\sqrt{N}$ with $N:=N_{+}+N_-$, we see from Eq.~\eqref{eq:delN} that the uncertainty in the measurement of the frequency $\Phi$ is not better than 
$1/\sqrt{N}\alpha T$.
Similarly, the statistical uncertainty in the EDM due to neutron counting is estimated as ${1}/{2\alpha E_0 T\sqrt{N}}$.
According to Ref.~\cite{Pendlebury:2015lrz}, combining the applied electric field of $E_0=7$ kV/cm, the averaged polarization $\alpha=0.58$, $T=130$ s and $N=2.5\times 10^9$, and also removing from consideration the measurement cycles with $E_0=0$ kV/cm, the statistical uncertainty is estimated as $1.34\times 10^{-26}\,e$cm, which is consistent with a more sophisticated estimation at a few percent level.
Similarly, the statistical uncertainty of $\Phi T$ is also estimated as $\sigma_{\Phi T}\simeq 3.71\times 10^{-5}$.
The point is that the phase of the cosine function of Eq.~\eqref{eq:Npm} is kept under control within the estimated uncertainty of $\sigma_{\Phi T}$ in the experiment~\cite{Pendlebury:2015lrz}.
%

\begin{figure}[t]
\begin{center}
\includegraphics[width=8.5cm]{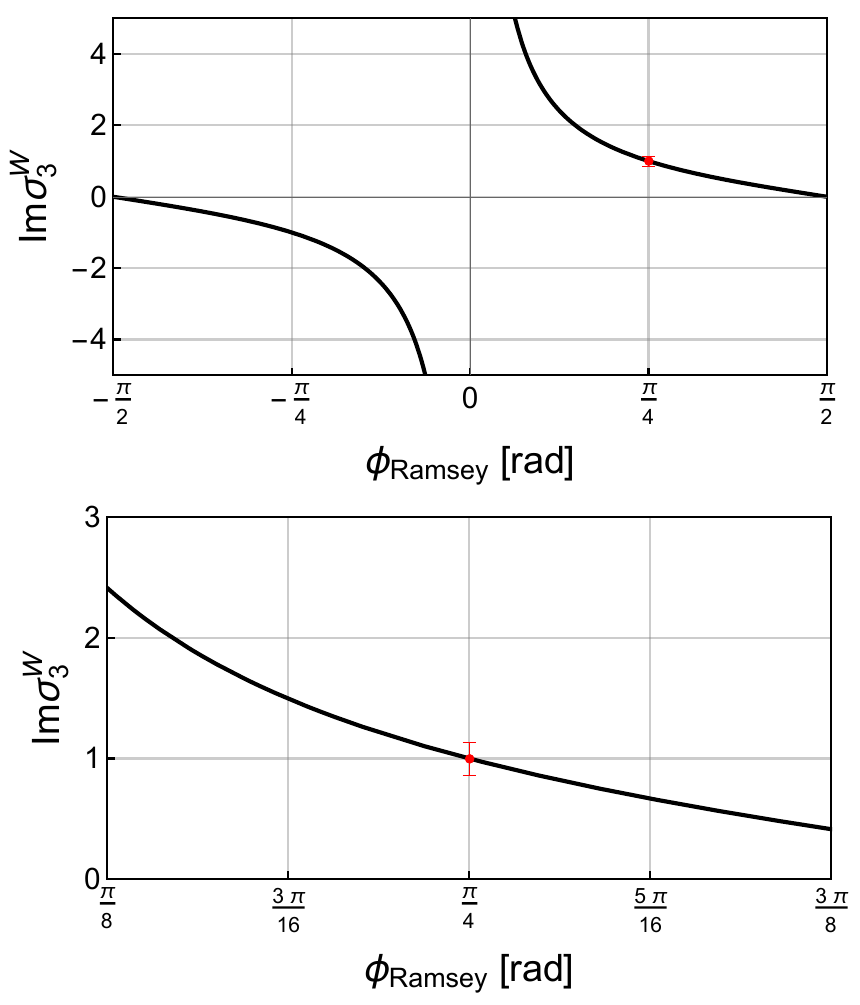}
\end{center}
\vspace{-0.4cm}
\caption{The imaginary component of the weak value of $\sigma_3$ of Eq.~\eqref{eq:WEAKNEU} (black line) for various $\phi_{\rm Ramsey}$, together with a past experimental value~\cite{sponar2015weak} (red point).
The upper plot is shown in a range $[-\pi/2,\pi/2]$, while the bottom depicts the enlarged plot focused in the range $[\pi/8, 3\pi/8]$ with an error bar given by the uncertainty. 
}.
\label{fig:dis}
\end{figure}
\begin{figure}[t]
\begin{center}
\includegraphics[width=8.5cm]{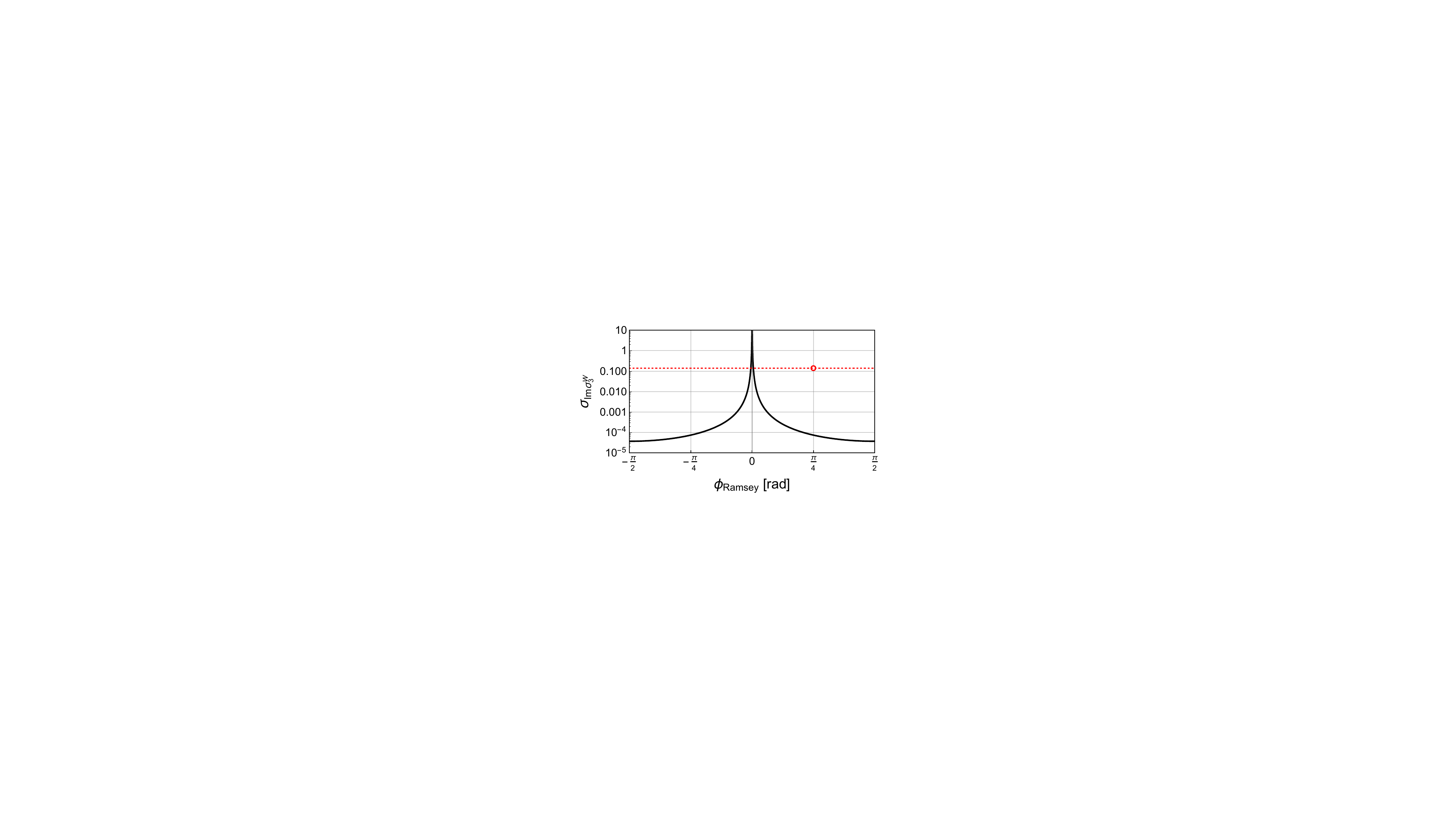}
\end{center}
\vspace{-0.4cm}
\caption{The uncertainty of the imaginary component of the weak value of $\sigma_3$ of Eq.~\eqref{eq:weakun} (black line), together with a past experimental value~\cite{sponar2015weak} (dotted red line), which corresponds to the error bar of Fig.~\ref{fig:dis}.
We assumed $\sigma_{\Phi T}\simeq 3.71\times 10^{-5}$ in the whole range.
}.
\label{fig:dis2}
\end{figure}

Now, let us return to the weak value of Eq.~\eqref{eq:weakRams}.
Using Eqs.~\eqref{eq:runN}, \eqref{eq:cycleN} and \eqref{eq:weakRams}, for each measurement cycle $j$, the imaginary part of the weak value $\sigma_3^W$ is obtained as 
\begin{align}
    ({\rm Im}\,\sigma^W_3)_j&=\frac{1}{\epsilon_{\pm,j} T}\frac{\cos\left(\delta \omega -\Phi_{\pm,j}\right)T-\cos\left(\delta \omega -\Phi_{\pm}\right)T}{1-\cos\left(\delta \omega -\Phi_{\pm,j}\right)T}\notag
    \\
    &\simeq \frac{\sin\left(\delta \omega -\Phi_{\pm}\right)T}{1-\cos\left(\delta \omega -\Phi_{\pm}\right)T}\notag
    \\
    &=\cot \phi_{\rm Ramsey}
    ,\label{eq:WEAKNEU}
\end{align}
where in the second line we omitted the first order of the EDM effects.
%
%
%
Fig.~\ref{fig:dis} shows the values of ${\rm Im}\,\sigma^W_3$ as a function of $\phi_{\rm Ramsey}$.
The neutron EDM experiments determine the phases of trigonometric functions of the right-hand side of Eq.~\eqref{eq:WEAKNEU}, which in turn determines the weak value with the uncertainty,
\begin{align}
    \sigma_{{\rm Im}\,\sigma^W_3}:&=\left|\csc\left(\frac{\left(\delta \omega -\Phi_{\pm}\right)T}{2}\right)\right|^2\sigma_{\Phi T}\notag
    \\
    &=\left|\csc\phi_{\rm Ramsey}\right|^2\sigma_{\Phi T},
    \label{eq:weakun}
\end{align}
as shown in Fig.~\ref{fig:dis2} as a function of $\phi_{\rm Ramsey}$.
For $\phi_{\rm Ramsey}=\pm \pi/4$ at which the red point of Fig.~\ref{fig:dis2} resides, we have the theoretical uncertainty $\sigma_{{\rm Im}\,\sigma^W_3}\simeq 7.42\times 10^{-5}$ under the reference value $\sigma_{\Phi T}$ mentioned before, which is far below the uncertainty in the preceding experiment of Ref.~\cite{sponar2015weak}.

\section{Discussions}
\label{sec:diss}
Before closing, we will discuss the feasibility of our method analyzed theoretically above in the light of the actual weak value measurement conducted earlier for the neutron's spin.
The weak value of the Pauli spin operator $\sigma_3$ of neutrons has been previously measured at the high-flux reactor of the Institute Laue-Langevin (ILL)~\cite{sponar2015weak} in Grenoble, France, by using the neutron interferometer and neutron beams, where the real and imaginary components of the weak value,
\begin{align}
    {\langle \sigma_3 \rangle}_{\rm w}:&=\frac{{\langle \Psi_f|\sigma_3|\Psi_i\rangle}}{{\langle \Psi_f|\Psi_i\rangle}},
\end{align}
has been reported.  There, the pre- and post-selected states are chosen as 
${|\Psi_i\rangle}= e^{-i\pi \sigma_2/4}{|+\rangle}$ and ${|\Psi_f\rangle}= e^{-i\alpha \sigma_3/2}e^{-i\beta \sigma_2/2}{|+\rangle}$.
In particular, for $\beta=-\pi/2$, we obtain
\begin{align}
    {\langle \Psi_f|\Psi_i\rangle}&={\langle +|e^{-i\pi\sigma_1/4}e^{i\alpha \sigma_3/2}e^{-i\pi\sigma_1/4}|+\rangle},\label{eq:nuPA}
    \\
    {\langle \Psi_f|\sigma_3|\Psi_i\rangle}
    &={\langle +|e^{-i\pi \sigma_1/4}e^{i\alpha \sigma_3/2} \sigma_3e^{-i\pi \sigma_1/4}|+\rangle}.\label{eq:dePA}
\end{align}
%
Recall that, for ${|\psi_i(\phi_{\rm Ramsey})\rangle}= {|\psi_f(\phi_{\rm Ramsey})\rangle}= e^{-i\phi_{\rm Ramsey}\sigma_2/2}{|+\rangle}$,  our weak value \eqref{eq:weak3} becomes
\begin{align}
    \sigma^W_3= \frac{{\langle \psi_f(\phi_{\rm Ramsey})|e^{-i\omega_2\tau\sigma_1}\sigma_2|\psi_i(\phi_{\rm Ramsey})\rangle}}{{\langle \psi_f(\phi_{\rm Ramsey})|e^{-i\omega_2\tau\sigma_1}|\psi_i(\phi_{\rm Ramsey})\rangle}}.\notag
\end{align}
For $\omega_2 \tau=\pi/2$, the numerator and denominator are found to be
\begin{widetext}
\begin{align}
    {\langle \psi_f(\phi_{\rm Ramsey})|e^{-i\omega_2\tau\sigma_1}|\psi_i(\phi_{\rm Ramsey}\rangle}&={\langle +|e^{-i\pi \sigma_1/4}e^{-i \phi_{\rm Ramsey}\sigma_3}e^{-i\pi \sigma_1/4}|+\rangle},\label{eq:nuOU}
    \\
    {\langle \psi_f(\phi_{\rm Ramsey})|e^{-i\omega_2 \tau \sigma_1}\sigma_2|\psi_i(\phi_{\rm Ramsey})\rangle}&={\langle +|e^{-i\pi \sigma_1/4} e^{-i\phi_{\rm Ramsey} \sigma_3}\sigma_3 e^{-i\pi\sigma_1/4}|+\rangle},\label{eq:deOU}
\end{align}
\end{widetext}
with $\phi_{\rm Ramsey}$ given in \eqref{eq:Ramseyphase}.
Observe then that, for $\alpha=-2\phi_{\rm Ramsey}$, Eqs.~\eqref{eq:nuOU} and \eqref{eq:deOU} are equal to Eqs.~\eqref{eq:nuPA} and \eqref{eq:dePA}, respectively. 
Thus, for $\alpha=-2\phi_{\rm Ramsey}$, $\beta=-\pi/2$, and $\omega_2\tau=\pi/2$, ${\langle \sigma_3\rangle}_{\rm w}=\sigma^W_3$ holds.
In Ref.~\cite{sponar2015weak}, ${\rm Im}\,{\langle \sigma_3\rangle}_{\rm w}$ was measured for $\alpha=-\pi/2$ and $\beta=-\pi/2$\footnote{
Unfortunately, no other measured points in Ref.~\cite{sponar2015weak} correspond to the fixed condition $\omega_2\tau =\pi/2$ in the neutron EDM experiment.
} as shown in Fig.~\ref{fig:dis} and \ref{fig:dis2}.
From Fig.~\ref{fig:dis2}, it is clear that the weak value measurement via the Ramsey resonance may improve the uncertainty of the imaginary component of the neutron's $\sigma_3$ by three orders of magnitude compared to Ref.~\cite{sponar2015weak}.
Only a single value in Ref.~\cite{sponar2015weak} can be compared here because the neutron EDM experiments take values in limited parameter ranges ($\omega_2\tau=\pi/2$ and $\omega\simeq \omega_0\pm \pi/2T$) to maximize its sensitivity to the neutron EDM.

However, it should be emphasized that, while neutron EDM experiments may be suited to precisely measure the imaginary part of $\sigma_3$, the conventional neutron beam experiments can measure both the real and imaginary parts of weak values.  
Besides, the conventional experiments using the neutron interferometer are indirect measurements and might be more suited for physical realization of quantum paradoxes such as the quantum Cheshire Cat~\cite{Denkmayr2014}.

\section{Summary}
\label{sec:5}

In the present paper we revisited the two typical resonance phenomena, the Rabi resonance and the Ramsey resonance, and showed that their response to deviation from the resonance point can be interpreted as the perturbative effect (represented by the term $V$ in the Hamiltonian) in the weak measurement.
Our main results given in Sec.~\ref{sec:3} are:
(i) Both of the Rabi and Ramsey resonances when the significant enhancement of the transition probabilities amount to the weak value amplification.
(ii) Near the resonance point, the transition probability becomes identical in form for both the Rabi and Ramsey resonances differing only in their measurement strengths.
(iii) The differences between the two resonances in the measurement strength and in the sensitivity to disturbance can be understood based on the probability formula shared by them.
In short, we found that the viewpoint of weak measurement provides a unified understanding of the Rabi and Ramsey resonances, where the the resonance phenomena are characterized by the behavior of the measurement strength under external conditions.

We also argued that by exploiting the data of previous neutron EDM experiments one can determine the imaginary component of the weak value of the spin of the neutrons.  Moreover, we disscussed that the precision of the obtained weak value turns out to be much higher (by three orders of magnitude) than that of the conventional weak value measurements using neutron beams.  Although the conventional method has its merit in being capable of measuring  
the real and imaginary parts of weak values, our observation indicates that novel advantage may be uncovered by shedding a new light on resonances from the viewpoint of weak value amplification.   
We hope that experimentalists conduct experiments following the methodology outlined in this paper.

We end with a few remarks on the outlook of this work.
First, weak value measurements through ground-state hyperfine transition frequency of the cesium 133 atoms may be possible because they are also based on the Ramsey resonance.  
It would be interesting there to investigate the weak values associated with the minimal disturbance of the SI time definition.
Second, the fundamental parameter measurements via the resonance might lead to a technology that allows one to realize quantum paradoxes such as the quantum Cheshire Cat~\cite{Denkmayr2014,aharonov2021dynamical,aharonov2024angular,Ghoshal:2022bnc} and the Three Box Paradox~\cite{Aharonov1991CompleteDO}, which involve the intriguing aspect of peculiar quantum existence and have been discussed primarily in the context of weak measurement.  
Third, it could be possible to find the counterpart of resonance in indirect measurement through the connection between weak value amplification and resonances mentioned here.  In fact, since all the weak measurements discussed in this paper belong to the framework of direct measurements discussed earlier~\cite{PhysRevLett.111.023604,PhysRevA.101.042117,RevModPhys.86.307}, we may expect more versatile outcomes if we adopt our argument in the framework of indirect measurements commonly used for the weak measurement.
On the other hand, the discussion of the merit on the side of spectroscopy is missing at present, and this will certainly be an important issue of future study for promoting practical application of weak measurement coupled with quantum resonance.  

\section*{Acknowledgments}
We sincerely thank Yuji Hasegawa and Takashi Higuchi for valuable discussions about experiments on neutron EDM.
This work was supported by JSPS KAKENHI Grant Number 20H01906.

\appendix

\bibliography{ref}

\end{document}